\documentclass[draft]{agujournal2019}
\usepackage{url} 
\usepackage{lineno}
\usepackage[inline]{trackchanges} 
\usepackage{soul}

\usepackage{CJK}
\usepackage{mathtools}
\usepackage{amsthm,amsmath,amssymb,lipsum}
\usepackage{mathrsfs}
\usepackage{cases}
\usepackage{algorithm}
\usepackage{algorithmicx}
\usepackage{algpseudocode}
\usepackage[inline]{trackchanges} 
\usepackage{soul}
\usepackage{threeparttable}
\usepackage{ragged2e}
\allowdisplaybreaks[4]
\usepackage{graphicx}
\usepackage{float}
\usepackage[T1]{fontenc}
\usepackage[utf8]{inputenc}
\usepackage{tabularx}
\usepackage{booktabs}
\usepackage{multirow}
\usepackage{multicol}
\usepackage{indentfirst}
\usepackage{xcolor}
\usepackage[a4paper]{geometry}
\usepackage{lmodern}
\usepackage{makecell}
\usepackage{array}
\usepackage{caption}

\draftfalse

\journalname{Journal of Geophysical Research: Machine Learning and Computation}

\begin{document}
\begin{sloppypar}
\newtheorem{definition}{Definition}
\newtheorem{theorem}{Theorem}[section]
\newtheorem{lemma}{Lemma}[section]

\title{Optimization Algorithm for Determining the Source Surface Radius Based on Parker Solar Probe in situ Measurements from Encounters 1 to 19}

\authors{Shiouhe Wang\affil{1,2},Fang Shen$^{*}$\affil{1,2},Yi Yang\affil{1,2},Xueshang Feng\affil{1,4},Jiansen He\affil{3}}

\affiliation{1}{SIGMA Weather Group,State Key Laboratory of Solar Activity and Space Weather,National Space Science Center,Chinese Academy of Sciences,Beijing 100190, People's Republic of China}
\affiliation{2}{College of Earth and Planetary Sciences,University of Chinese Academy of Sciences,Beijing 100049, People's Republic of China}
\affiliation{3}{School of Earth and Space Sciences, Peking University, 100871 Beijing, People's Republic of China}
\affiliation{4}{Shenzhen Key Laboratory of Numerical Prediction for Space Storm, School of Aerospace, Harbin Institute of Technology, Shenzhen 518055, People's Republic of China}

\correspondingauthor{Fang Shen}{fshen@spaceweather.ac.cn}

\begin{keypoints}
\item We prove the well-posedness for PFSS extrapolation and show its inverse problem admits a global minimum via compactness and continuity.
\item We construct an optimization algorithm to estimate $R_{ss}$ from comparisons between PFSS extrapolation and PSP in situ measurements.
\item ACE validation and Pareto analysis show the optimal $R_{ss}$ generally increases with solar activity while balancing open flux and polarity.

\end{keypoints} 

\begin{abstract}
The Potential Field Source Surface (PFSS) extrapolation is a method for estimating the large scale coronal magnetic field from photospheric magnetograms. 
The source surface serves as the outer boundary of its solution domain, 
and is typically a spherical surface. 
An appropriate source surface radius ($R_{ss}$) enables more accurate identification of the coronal magnetic field topology and estimation of the open flux, 
thereby potentially enhancing the accuracy of space weather modeling.
We prove the well-posedness of the PFSS forward problem and establish the existence and uniqueness of the optimal source surface by combining compactness of the admissible set with continuity of the objective functional. 
The objective functional is the mean squared error (MSE) between PFSS extrapolation and Parker Solar Probe (PSP) radial magnetic field measurements after Parker spiral backmapping and radial scaling for Encounters 1-19. 
The optimization algorithm is validated with an analytical solution, 
and Advanced Composition Explorer (ACE) in situ measurements are used as an independent cross-validation dataset. 
Additional evaluation metrics and Pareto analysis are used to identify the dominant metrics between open flux and polarity prediction accuracy. 
Our results show that the optimal $R_{ss}$ derived from the algorithm generally increase from solar minimum into the ascending phase of solar cycle 25. 
The optimized solution improves open flux agreement while preserving or improving polarity prediction accuracy relative to $2.5R_{s}$. 
The Pareto frontiers show a transition for dominant metrics from open flux during solar minimum to polarity prediction accuracy during the ascending phase.
\end{abstract}

\section*{Plain Language Summary}
The Potential Field Source Surface extrapolation estimates the Sun's large scale coronal magnetic field from photospheric magnetic maps. 
Its geometry parameter $R_{ss}$ is the source surface radius, 
and affects which magnetic field lines are open to the heliosphere and how the interplanetary magnetic field in situ measurements backmaps to the solar surface or low corona.
In the present study, 
we use radial magnetic field and solar wind velocity from Parker Solar Probe (PSP) measurements to construct an optimization algorithm to identify which $R_{ss}$ parameter is the optimal value relative to the observations. 
We then compare the optimal $R_{ss}$ derived from the algorithm and its trend with independent Advanced Composition Explorer (ACE) in situ measurements near 1 Astronomical Unit. 
The optimal $R_{ss}$ generally increases from solar minimum into the ascending phase of solar cycle 25. 
The optimized solution improves agreement with open flux calculations while preserving or improving the predicted magnetic field polarity compared to the conventional $2.5R_{s}$, 
which helps clarify how to choose $R_{ss}$ under different solar activity cycles.

\section{Introduction} \label{sec:introduction}

The Potential Field Source Surface (PFSS) extrapolation estimates the large scale coronal magnetic field from a photospheric magnetogram by solving Laplace equation for the magnetic scalar potential in the domain between the photosphere and a prescribed source surface.
In this extrapolation, the source surface serves as the outer geometric boundary, where the coronal magnetic field is assumed to become radial.
The source surface radius ($R_{ss}$) can influence the modeled distribution of open and closed magnetic field lines, the estimated open magnetic flux, and the candidate solar source regions inferred from backmapping interplanetary in situ measurements.
\citeA{schatten_model_1969} estimates $R_{ss}$ at $1.6$ solar radius ($R_s$) by comparing PFSS extrapolations with interplanetary observations.
\citeA{altschuler_magnetic_1969} suggests $R_{ss}=2.5R_s$ by comparing magnetic field line structures from PFSS extrapolations with white light coronal images during the ascending phase of the solar cycle.
Subsequent analyses of heliospheric current sheet (HCS) topology and interplanetary magnetic field (IMF) polarity also suggest $R_{ss}=2.5R_s$ \cite{https://doi.org/10.1029/JA088iA12p09910}.
The value $R_{ss}=2.5R_s$ has therefore become a conventional reference choice in PFSS extrapolation studies, 
with applications to estimating the flux tube expansion factor in the Wang-Sheeley-Arge (WSA) model 
\cite{arge_improved_2003,https://doi.org/10.1029/2004JA010384,SHEN2012125,Yalim_2017,Liu_2019,Dakeyo_2024b}, 
initializing coronal magnetohydrodynamic (MHD) simulations 
\cite{https://doi.org/10.1029/2006JA012164,SHEN20101008,https://doi.org/10.1029/2011JA016584,https://doi.org/10.1029/2010JA015809,Shen_2018}, 
studying solar energetic particle transport 
\cite{Zhao_2018,WEI2019155,universe10080315,Park_2024}, 
and evaluating coronal and heliospheric magnetic field models with observations 
\cite{Lowder_2014,https://doi.org/10.1002/2015JA021757,Virtanen_2022}.
\citeA{Zhao_2025} uses $R_{ss}=2.5R_s$ in a statistical analysis of IMF evolution over heliocentric distances from 0.1 to 30 astronomical units (AU).

Despite its widespread applications, $R_{ss}=2.5R_s$ can lead to discrepancies 
in the modeled extent of coronal holes and the estimation of the open magnetic flux, 
indicating that $R_{ss}$ is a tunable parameter and the optimal $R_{ss}$ need be determined by observational constraints. 
For instance, \citeA{lee_coronal_2011} analyzes the extent of coronal holes during the minimum phases 
of solar cycles 22 and 23 to estimate the optimal $R_{ss}=1.9R_s$ and $1.8R_s$, respectively.
\citeA{https://doi.org/10.1002/2013JA019464} compares the modeled open flux with observations near 1 AU 
and shows that a larger $R_{ss}$ is required during the maximum phase of solar cycle 23.
A lower $R_{ss}$ ($1.5R_{s}\sim 2.0R_{s}$) is identified during the solar cycle 24 through comparisons between PFSS extrapolations and IMF measurements near 1 AU \cite{https://doi.org/10.1029/2019SW002205}.
Total Solar Eclipse (TSE) white light images serve as an essential observable source for coronal magnetic field structures and are used to identify the optimal $R_{ss}$ \cite{Boe_2020,Habbal_2021}.
\citeA{Benavitz_2024} analyzes the coronal magnetic field topology during the solar cycle 24 maximum and minimum phases using TSE images,
suggesting that the optimal $R_{ss}$ is approximately $1.3R_s$ during the maximum phase of solar cycle and $3.0R_s$ during the minimum phase.
\citeA{Huang_2024} compares the extent of open magnetic field regions in PFSS and MHD models and concludes that $R_{ss}$ increases with solar activity.
Further constraints from PFSS extrapolations and in situ measurements also point to a broad range of $R_{ss}$ spanning $1.3R_s$ to $3.5R_s$ \cite{Koskela_2018,Koskela_2019,Virtanen_2019,https://doi.org/10.1029/2019JA027173,https://doi.org/10.1029/2020JA028870,Finley_2023,Ismo_2024}.

Parker Solar Probe (PSP) completed its first encounter during the minimum phase of solar cycle 25 and has since become the closest spacecraft to the Sun to date \cite{fox_solar_2016,raouafi_parker_2023}.
PSP provides an unprecedented opportunity to constrain $R_{ss}$ through comparisons between PFSS extrapolations and in situ measurements near $10R_{s}$.
PFSS extrapolation is commonly used to trace the source regions of solar wind because its computational efficiency and high spatial resolution makes repeated magnetic field lines tracing practical \cite{schrijver_nonlinear_2006,Badman_2020}.
The source region mapping is traced through a two-step procedure. 
The spacecraft position is first ballistically mapped back to source surface footpoints along Parker spiral trajectory using the radial solar wind velocity of in situ measurements \cite{Parker1958,Nolte1973}. 
Magnetic field lines from the PFSS extrapolation are then traced from these footpoints to the photosphere or low corona.
Several studies using PSP in situ measurements have shown that the candidate solar source regions inferred by ballistic backmapping and subsequent magnetic field tracing with PFSS extrapolations can vary substantially with the adopted $R_{ss}$ \cite{Dakeyo_2024,Koukras_2025,Bizien_2025}.
For slowly Alfv\'enic solar wind observed during PSP Encounters 4-14, 
the footpoint locations of solar wind are found to be most sensitive to the adopted $R_{ss}$ \cite{TamarErvin_2024}.
In a source region analysis of slow Alfv\'enic solar wind observed near low latitudes, 
a two-step ballistic backmapping method with $R_{ss}=2.0R_s$ was used to trace the large scale magnetic structure of the solar wind \cite{bale_highly_2019}.
\citeA{Badman_2020} suggests that using a lower $R_{ss}$ ($1.3R_{s}\sim 1.5R_{s}$) improves polarity inversion predictions based on PSP in situ measurements during Encounter 1.
\citeA{de_pablos_searching_2022} adopts $R_{ss}=2.0R_s$ to trace the source regions of polarity inversions during PSP Encounter 1.
\citeA{Bercic_2020} uses this backmapping procedure with $R_{ss}=2.0R_s$ and indicates coronal electron temperatures consistent with Strahl electron observations from PSP and Helios in the inner heliosphere.
\citeA{hou_origin_2024} traces switchbacks observed from PSP Encounter 4 back to candidate solar source regions by using a two-step ballistic backmapping procedure with $R_{ss}=2.0R_s$, 
and suggests that these switchbacks originate from magnetic reconnection at chromospheric network boundaries.

In summary, previous studies suggest that no single $R_{ss}$ is universally preferred.
\citeA{Badman_2022} shows that PFSS extrapolations do not generally yield a single globally optimal $R_{ss}$ when coronal hole distribution, streamer belt topology, and magnetic polarity are evaluated simultaneously. 
This result suggests that the preferred $R_{ss}$ depends on which observational constraints and evaluation metrics are emphasized. 
To characterize such trade-offs, we use a Pareto analysis, in which a solution is regarded as Pareto optimal if no objective can be improved without degrading at least one other objective \cite{zbMATH01349589,zbMATH01614566,CHANG20151105}.
It remains unclear whether the optimization of $R_{ss}$ exhibits a well-defined Pareto frontier among competing evaluation metrics, 
and how the optimal $R_{ss}$ varies with solar activity for a prescribed objective functional.

\begin{figure}[ht!]
      \includegraphics[width=35pc]{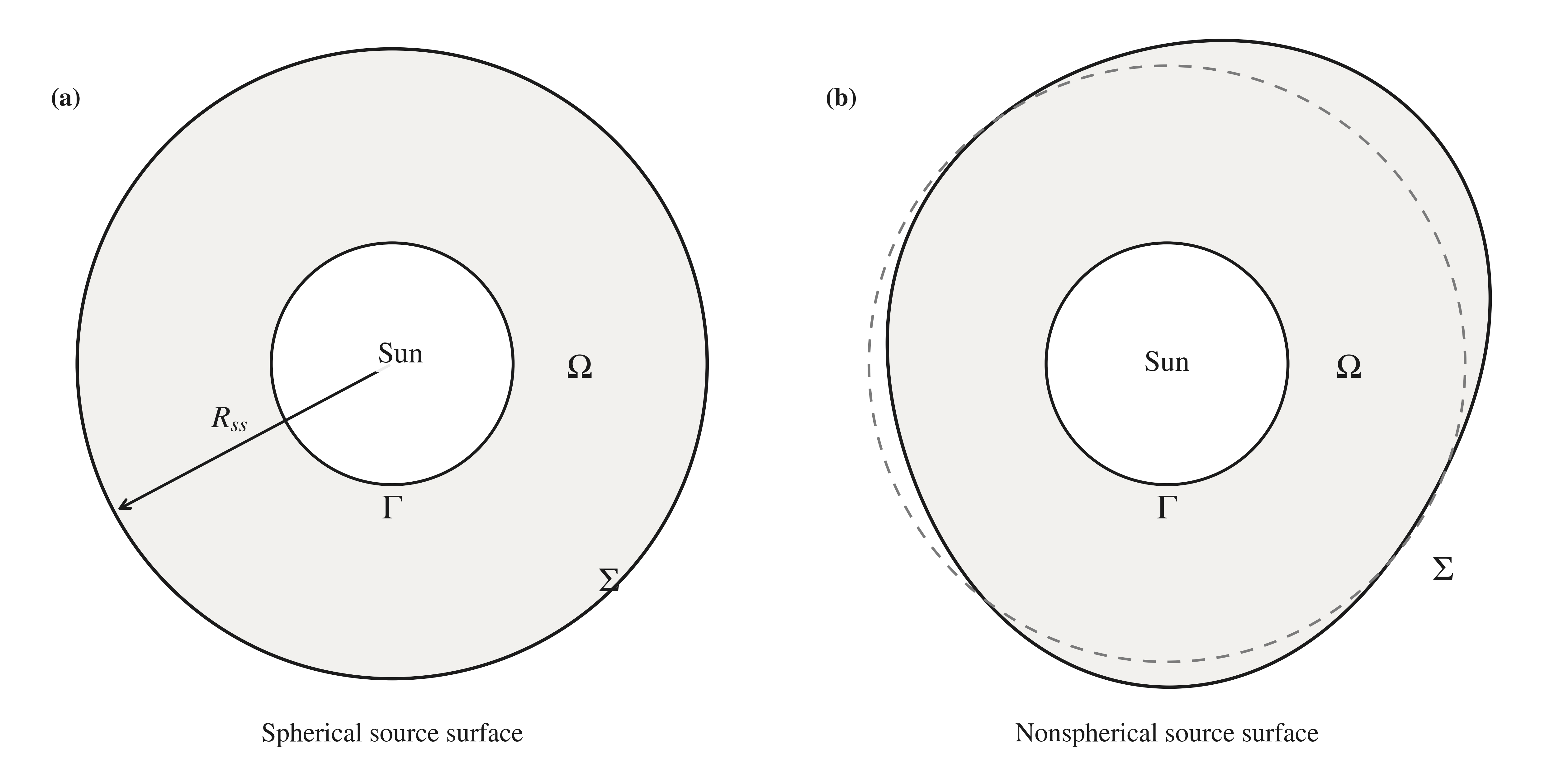}
      \caption{(a) Illustration of the solution domain for PFSS extrapolation with a spherical source surface of radius $R_{ss}$. 
      (b) Illustration of the solution domain for PFSS extrapolation with a nonspherical source surface, 
      where $\Omega$ denotes the solution domain, 
      $\Gamma$ denotes the inner fixed boundary,
      and $\Sigma$ denotes the outer free boundary.}
      \label{fig:Geometry}
\end{figure}
Previous studies have explored nonspherical source surface as a way to improve the agreement between coronal magnetic field extrapolations and observations \cite{schulz_coronal_1978,levine_simulation_1982,angeo-15-1379-1997,Panasenco_2020,Kruse_2020_and_Heidrich,Kruse_2021}.
Figure \ref{fig:Geometry}(a) shows the spherical reference geometry, 
and Figure \ref{fig:Geometry}(b) shows a nonspherical geometry with the same inner fixed boundary but an outer free boundary.
For a prescribed outer boundary, 
such nonspherical boundary value problems can be solved numerically, 
for example by finite element methods (FEMs) \cite{Kang1996}. 
In many practical applications, the source surface geometry is not known a priori. 
Some studies have considered prescribed nonspherical geometries, 
such as an elliptic outer boundary \cite{Kruse_2020_and_Heidrich,Kruse_2021}. 
This geometry in advance would introduce an artificial restriction on the coronal field topology.
If additional observational condition is imposed on the outer boundary, 
the resulting boundary conditions will become overdetermined for a fixed domain. 
The natural mathematical formulation is then a free boundary or shape optimization problem, 
in which the magnetic field and the outer boundary geometry need be determined simultaneously. 
Shape optimization provides a rigorous framework for this problem by minimizing a prescribed objective functional over an admissible class of domains \cite{Sokolowski1992}. 
This motivates a standardized shape optimization formulation for nonspherical coronal magnetic field extrapolations.

Section~\ref{sec:2} establishes the well-posedness of the forward PFSS problem and formulates the estimation of $R_{ss}$ within a free boundary framework. 
We then prove the compactness of the admissible class and the continuity of the objective functional, 
which together imply the existence of a global minimum for this inverse problem. 
These arguments apply directly to nonspherical source surface geometries, and are not restricted to the spherical case.
Section \ref{sec:3} describes the magnetogram, PSP, and ACE datasets and clarifies how to integrate the PFSS extrapolation, Parker spiral backmapping, and cross-validation with these observations. 
Section \ref{sec:4} defines the objective functional, 
constructs the optimization algorithm and its validation, 
and introduces the additional evaluation metrics and Pareto analysis. 
Section \ref{sec:result} shows the optimal values of $R_{ss}$ for PSP Encounters 1-19, 
examines how $R_{ss}$ vary with solar activity, 
and analyzes the results using the evaluation metrics and Pareto frontiers.
Section \ref{sec:5} concludes the paper.

\section{Mathematical Formulation} \label{sec:2}

In this section, we formulate the source surface inverse problem as a free boundary problem. 
The formulation applies to both spherical and nonspherical source surfaces.
We first prove the well-posedness of the forward problem for PFSS extrapolation and then prove the existence and uniqueness of a minimizer for the associated free boundary optimization problem. 
For the forward problem, the well-posedness means the existence and uniqueness of the PDE solution $\psi$ for a prescribed domain. 
For the inverse problem, we need to establish the existence and uniqueness of a source surface geometry of minimizing the objective functional over the admissible set. 
The compactness and continuity of the admissible set and the objective functional need to be established.
These arguments in this section can apply directly to nonspherical source surface geometries, 
not only to the spherical case used later in the optimization algorithm.

\subsection{Description} \label{subsec:2.0}
The forward problem for PFSS extrapolation is a second order elliptic boundary value problem with mixed boundary conditions and therefore belongs to a class of steady state problems.
If both its solution and the outer boundary need to be determined from additional boundary conditions or observational constraints, 
the problem is formulated as a free boundary problem \cite{Tepper1975,acker_interior_1981}. 
Due to the internal Neumann boundary condition for PFSS extrapolation in the forward problem, 
existing research results for free boundary problems cannot be applied directly, 
so the proof must be rebuilt for the setting of the PFSS extrapolation.

We first introduce the functional definitions used in the variational formulation for the free boundary problem. 
Since the equation underlying the PFSS extrapolation is an elliptic boundary value problem for the scalar potential $\psi$, 
and the magnetic field is given by $\mathbf{B}=-\nabla\psi$, the natural energy space is $H^1(\Omega)$. 
This space requires only square integrable weak derivatives, gives finite magnetic energy 
$\int_{\Omega}|\nabla\psi|^2\,dx$, 
and provides well-defined boundary traces for imposing Dirichlet and Neumann conditions. 
More mathematical foundations of Sobolev spaces and elliptic PDEs can be found in classical references \cite{strauss2008pde,evans2010pde}.

Let $m\in\mathbb{N}$ and $p\in[1,\infty]$. 
The Sobolev space $W^{m,p}(\Omega)$ is defined as
\begin{equation}
     W^{m,p}(\Omega)\coloneq
     \left\{
     u\in L^{p}(\Omega)\ \middle|\ 
     \partial^{\alpha}u\in L^{p}(\Omega),\ 
     \forall \alpha\in \mathbb{N}^3,\ |\alpha|\leq m
     \right\},
     \label{con:Eq.4}
\end{equation}
where $\alpha$ is a multi-index and $L^{p}(\Omega)$ denotes the Lebesgue $p$-integrable functions space.
For $1\leq p<\infty$, $W^{m,p}(\Omega)$ is equipped with the norm
\begin{equation}
     \Vert u \Vert_{W^{m,p}(\Omega)}
     \coloneq
     \left(
     \sum_{|\alpha|\leq m}
     \Vert\partial^{\alpha}u\Vert_{L^{p}(\Omega)}^{p}
     \right)^{1/p}.
     \label{con:Eq.5}
\end{equation}
In particular, $W^{m,2}(\Omega)$ is denoted by $H^{m}(\Omega)$. 
In this work, the scalar potential in the PFSS extrapolation is sought in $H^1(\Omega)$, 
which is the standard space for the weak formulation of second order elliptic problems.
We next define the admissible set of domains for the free boundary problem.

\begin{definition}[The Admissible Set]
    Let $\Omega$ be the annular domain in $\mathbb{R}^{3}$ with the inner fixed boundary $\Gamma$ and the outer free boundary $\Sigma$.
    $\Gamma$ and $\Sigma$ are $C^{2}$-smooth and homeomorphic to the unit sphere.
    The admissible set $\mathcal{A}$ is
    \begin{equation}
    \mathcal{A} := \left\{ \Omega \subset \mathbb{R}^3 \,\middle|\, \partial\Omega = \Gamma \cup \Sigma,\; \Gamma \text{ is fixed and } \Sigma \text{ is free, both of class } C^2 \text{ and simply connected}\right\}.
    \end{equation}
    \label{def admissible set}
\end{definition}
The solution to the free boundary problem consists of finding a domain $\Omega \in \mathcal{A}$ that satisfies the additional boundary conditions, 
which is imposed on the free boundary $\Sigma$. 
The problem is formally defined as follows.

\begin{definition}[Free Boundary Problem]
    Let $\Omega \in \mathcal{A}$ be an annular domain, where $\mathcal{A}$ is defined in Definition \ref{def admissible set}. 
    The free boundary problem consists of finding a domain $\Omega$ and a function $\psi\in H^{1}\left(\Omega\right)$ such that
    \begin{equation} 
        \begin{cases}
            -\mathcal{L} \psi = 0 & \text{in } \Omega, \\
            \partial_{\mathbf{n}} \psi = g & \text{on } \Gamma, \\
            \psi = 0 & \text{on } \Sigma, \\
            \partial_{\mathbf{n}} \psi = h & \text{on } \Sigma, \\
        \end{cases} 
        \label{con:Eq.1}
    \end{equation}
    where $\mathcal{L}$ is the prescribed differential operator and the exterior normal derivative is $\partial_{\mathbf{n}}\psi\coloneq \nabla \psi\cdot\mathbf{n}$,
    with $\mathbf{n}$ denoting the outward unit normal on $\partial \Omega$.
    The Neumann data are $g\in H^{-\frac{1}{2}}\left(\Gamma\right)$ on the inner boundary $\Gamma$ and $h\in H^{-\frac{1}{2}}\left(\Sigma\right)$ on the outer boundary $\Sigma$.
    The Dirichlet condition prescribes zero potential on $\Sigma$.

\label{def FBP}
\end{definition}

Relevant studies in free boundary problems primarily focus on the external Bernoulli free boundary problem with internal Dirichlet condition \cite{Alt1984,ITO2006126,IVANYSHYNYAMAN20172784},
originating from the ideal fluid dynamics \cite{CaffarelliAlt+1981+105+144}.
Although the existence and uniqueness of the solution for this internal Dirichlet boundary value are well-established \cite{LIEBERMAN1986422,Julius}, 
these properties for the internal Neumann condition have not been proven.
The free boundary problem for PFSS extrapolation differs from the Bernoulli free boundary problem because it imposes an internal Neumann condition, 
which is weaker than the internal Dirichlet condition used in the Bernoulli free boundary problem.
The well-posedness of the problem need to be rebuilt and can be proven by analyzing the variational equations in the shape optimization framework \cite{Sokolowski1992,kawohl_antoine_2019,Julius}.
We apply the shape optimization framework to formulate this free boundary problem.

\begin{definition}[Objective Functional]
    The objective functional $\mathcal{J}\left(\Omega\right)$ is constructed from the two auxiliary problems \eqref{con:Eq.2.1} and \eqref{con:Eq.2.2}:

     \begin{subequations}\label{con:Eq.2}
          \begin{numcases}{\mbox{find $\Omega\in\mathcal{A}$}~:}
          \text{find $\psi_{D}\in H^1\left(\Omega\right)$, such that:}  
          \begin{cases}
               -\Delta \psi_{D}=0 & \text{in $\Omega$,} \\
               \partial_{\mathbf{n}} \psi_{D}=g & \text{on $\Gamma$,} \\
               \psi_{D}=0 & \text{on $\Sigma$,} \\
               \end{cases}
               \label{con:Eq.2.1}\\
               \text{find $\psi_{R}\in H^1\left(\Omega\right)$, such that:}   \label{con:Eq.2.2}
          \begin{cases}
               -\Delta \psi_{R}=0 & \text{in $\Omega$,} \\
               \partial_{\mathbf{n}} \psi_{R}=g & \text{on $\Gamma$,} \\
               \partial_{\mathbf{n}} \psi_{R}+\beta \psi_{R} =h & \text{on $\Sigma$.} \\
          \end{cases} 
          \end{numcases}
     \end{subequations}
     where $\psi_{D}$ and $\psi_{R}$ are the two auxiliary solutions used to define the objective functional $\mathcal{J}\left(\Omega\right)$.

\label{def WFBP}
\end{definition}
The shape optimization minimizes the functional $\mathcal{J}$ over the admissible domains $\mathcal{A}$.
The well-posedness does not depend on the particular choice of objective functional.
In general, the single objective functional used in shape optimization is not unique, 
while the solution to the free boundary problem is invariant \cite{rabago_second-order_2020}.
Thus, the optimizer recovers the physically meaningful free boundary.

\begin{definition}[Shape Optimization]
     The shape optimization is formulated as a single objective optimization problem with PDE constraints.
    \begin{equation} 
     \begin{cases}
     \underset{\Omega\in\mathcal{A}}{\min}\mathcal{J}\left(\Omega\right)=\underset{\Omega\in\mathcal{A}}{\min}\int_{\Omega}\left|\nabla\left(\psi_{D}-\psi_{R}\right)\right|^{2}dx,\\
     F\left(\mathbf{x},\psi,\partial_{\mathbf{n}} \psi,\Delta\psi;\Omega\right)=0, \\
     \end{cases} 
     \label{con:Eq.3}
     \end{equation}
     where $F$ represents the combination of PDE and boundary conditions.
\label{def Shape Optimization}
\end{definition}
The shape optimization problem is equivalent to the free boundary problem.
We use variational calculus to prove the existence and uniqueness of Definition \ref{def Shape Optimization}, .
The key is to construct a solution set satisfying the PDE constraints and then study the property of functional $\mathcal{J}$ on this set.
If the set is compact and $\mathcal{J}$ is continuous, then $\mathcal{J}$ attains a minimum.
Continuity of $\mathcal{J}$ is straightforward; the main difficulty is to establish compactness.
We first need to prove the well-posedness of the forward problem for PFSS extrapolation and then analyze the compactness of the solution set.
Only the main steps are given here, and the detailed arguments are shown in \ref{sec:D}.

\subsection{The well-posedness of the forward problem for PFSS extrapolation} \label{subsec:2.1}
The well-posedness of the forward problem for PFSS extrapolation means the existence and uniqueness of the solution $\psi$.
This property is essential both for coronal magnetic field simulations and the free boundary problem. 
The forward problem for PFSS extrapolation is a mixed Dirichlet and Neumann boundary value problem, 
and we prove its well-posedness by reformulating it in variational form on the Sobolev space.
We first derive the variational equation and identify the corresponding solution space.
We then define the associated bilinear form and apply the Lax-Milgram theorem to prove the existence and uniqueness of the solution.
The proof is carried out for a general nonspherical source surface, with the spherical case included as a special case.

\begin{lemma}[\textbf{Variational Equation of the Forward Problem for PFSS Extrapolation}]
     A function $\psi \in V\left(\Omega\right)$ solves the forward problem for PFSS extrapolation if and only if for any $\phi \in V\left(\Omega\right)$,
     \begin{equation}
          \int_{\Omega}\left(\nabla \psi \right)\cdot \left(\nabla \phi \right) dx = \int_{\Gamma}g \gamma_{0}\left(\phi\right) d\sigma,
          \label{con:Eq.6}
     \end{equation}
     where $V\left(\Omega\right)=\left\{v \in H^{1}\left(\Omega\right): v|_{\Sigma}=0\right\}$ is equipped with the norm $\Vert\cdot\Vert_{H^{1}\left(\Omega\right)}$, 
     and $\gamma_{0}$ denotes the trace operator mapping functions to their boundary values.  
     
     \noindent \textbf{Proof.}
     Shown in \ref{sec:D}.
     \label{lemma:2.1}
\end{lemma}

This lemma establishes the equivalence between the forward problem for PFSS extrapolation and its variational formulation.
We apply the Lax-Milgram theorem to this variational equation and then yields the well-posedness of the forward problem.
We first define the bilinear functional $a\left(\cdot,\cdot\right)$: $V\times V\rightarrow \mathbb{R}$ by
\begin{equation}
     a\left(\psi,\phi\right)\coloneqq \int_{\Omega}\left(\nabla\psi\right)\cdot\left(\nabla\phi\right)dx,\quad \forall \left(\psi,\phi\right)\in V\times V,
     \label{con:Eq.15}
\end{equation}
and a linear functional $\eta_{g}$: $V\rightarrow \mathbb{R}$ by
\begin{equation}
     \eta_{g}\left(\phi\right)\coloneqq \int_{\Gamma}g\gamma_{0}\left(\phi \right)d\sigma,\quad \forall \phi \in V.
     \label{con:Eq.16}
\end{equation}
Then, the variational equation can be represented by these form:
\begin{equation}
     \eqref{con:Eq.6}\iff \text{Find } \psi \in V, \text{ such that } a\left(\psi,\phi\right)=\eta_{g}\left(\phi\right),\quad \forall \phi \in V.
     \label{con:Eq.17}
\end{equation}
If $a\left(\cdot,\cdot\right)$ and $\eta_{g}\left(\cdot\right)$ satisfy the assumptions of the Lax-Milgram theorem,
then the forward problem for PFSS extrapolation is well-posed.
We verify these assumptions in the following theorem.

\begin{theorem}[\textbf{Well-posedness of the Forward Problem for PFSS Extrapolation}]
The variational equation $a\left(\psi,\phi\right)=\eta_{g}\left(\phi\right)$ has a unique solution in $V$.

\noindent \textbf{Proof.}
Shown in \ref{sec:D}.
\label{thm:1}
\end{theorem}

The variational formulation for the forward problem extends naturally to the two auxiliary problems \eqref{con:Eq.2.1} and \eqref{con:Eq.2.2}.
Once a convergent sequence of the admissible domain leading to the shape optimization of \eqref{con:Eq.3} is constructed, 
the free boundary problem for PFSS extrapolation is well-posed.

\subsection{The well-posedness of the free boundary problem for PFSS extrapolation} \label{subsec:2.2}

We now define the solution space for the free boundary problem and prove its compactness.

\begin{definition}[Solution Space]
     To solve the optimization problem \eqref{con:Eq.3}, we define the space of triples
     \begin{equation} 
      \mathcal{M}\coloneq\left\{\left(\Omega,\psi_{D},\psi_{R}\right):\Omega\in\mathcal{A};\psi_{D},\psi_{R}\in H^{1}\left(\tilde{U}\right), \text{and satisfy the \eqref{con:Eq.2.1} and \eqref{con:Eq.2.2}}\right\},
     \end{equation}
     where $\tilde{U}$ is a fixed extension domain containing every admissible domain in $\mathcal{A}$.
\label{def Solution Set}
\end{definition}

Prior to analyzing the convergence of sequences within $\mathcal{M}$,
a parameterization of $\Omega$ is necessary. 
There exists a diffeomorphism $\varphi:\mathbb{S}^{2}\rightarrow \Sigma$ such that $\varphi$ uniquely determines the outer boundary $\Sigma\left(\varphi\right)$,
which in turn uniquely determines the admissible domain $\Omega\left(\Sigma\right)$. 
Each element in $\mathcal{A}$ can be characterized by $\varphi$.
We therefore define the convergence of domains $\Omega_{n}\rightarrow\Omega$ through the convergence of their parameterizations, $\varphi_{n}\rightarrow\varphi$.
\begin{figure}[ht!]
      \centering
      \includegraphics[width=35pc]{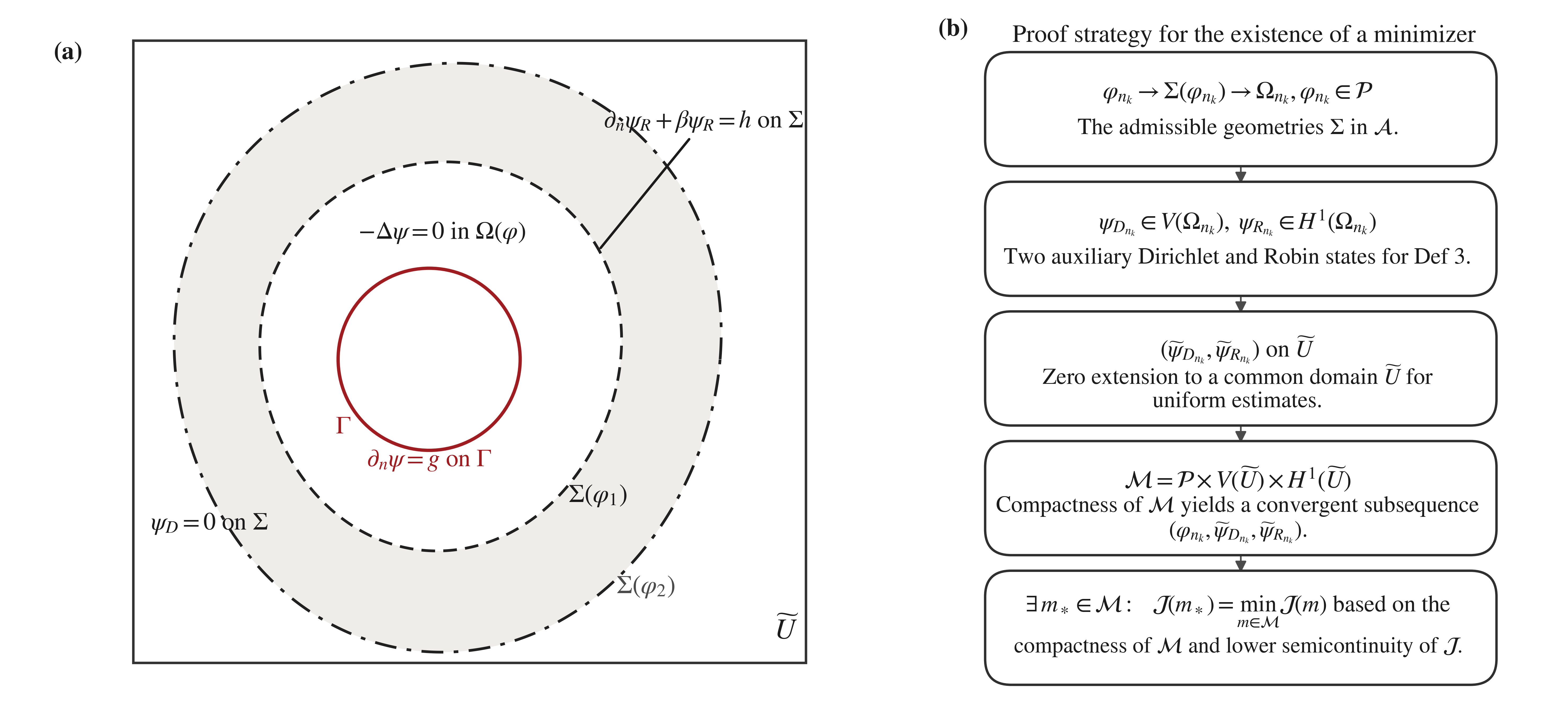}
      \caption{
      (a) Geometric setting of the free boundary problem for PFSS extrapolation.
      The two black dashed source surface boundaries represent admissible variations of the outer free boundary, 
      where red boundary$\Gamma$ denotes the inner fixed boundary and $\widetilde{U}$ denotes the fixed extension domain. 
      (b) Proof strategy of free boundary problem. 
      The diagram summarizes the passage from admissible geometries and the auxiliary Dirichlet and Robin states to their zero extensions on $\widetilde{U}$, 
      followed by the compactness and lower semicontinuity that yields a minimizer of the objective functional $\mathcal{J}$.
      }
      \label{fig:Commutative}
\end{figure}
Figure \ref{fig:Commutative}(a) shows the geometric setting of the free boundary problem and the admissible variations of the outer boundary $\Sigma$. 
Figure \ref{fig:Commutative}(b) summarizes the proof strategy on the fixed extension domain $\widetilde{U}$ and shows how the auxiliary solutions \eqref{con:Eq.2.1} and \eqref{con:Eq.2.2} enter the compactness and lower semicontinuity that yields a common limit and a minimizer of the objective functional $\mathcal{J}$.
The following result establishes the compactness of $\mathcal{M}$.
This argument extends the Bernoulli free boundary framework to the free boundary problem for PFSS extrapolation and corrects the erroneous of proof given in \cite{Julius}.

\begin{lemma}[$\textbf{The Compactness of }\mathcal{M}$]
     Let $\left(\varphi_{n},\psi_{D_{n}},\psi_{R_{n}}\right)$ be a sequence in $\mathcal{M}$,
     where $\psi_{D_{n}}$ and $\psi_{R_{n}}$ are weak solutions of \eqref{con:Eq.2.1} and \eqref{con:Eq.2.2} in $\Omega_{n}=\Omega\left(\left(\Sigma\left(\varphi_{n}\right)\right)\right)$, respectively.
     Then, there exists a subsequence $\left(\varphi_{n_{k}},\psi_{D_{n_{k}}},\psi_{R_{n_{k}}}\right)$, a mapping $\varphi\in\mathcal{P}$ and functions $\psi_{D},\psi_{R}\in V\left(\tilde{U}\right),H^{1}\left(\tilde{U}\right)$, respectively, such that:
     \begin{equation} 
          \begin{cases}
               \varphi_{n_{k}}\rightarrow \varphi & \text{in $\mathcal{P}$,}\\
               \psi_{D_{n_{k}}}\rightarrow \psi_{D} & \text{in $V\left(\tilde{U}\right)$,}\\
               \psi_{R_{n_{k}}}\rightarrow \psi_{R} & \text{in $H^1\left(\tilde{U}\right)$,}\\
               \end{cases}
          \label{con:Eq.23}
     \end{equation}
     where $\psi_{D}$,$\psi_{R}$ are solutions to \eqref{con:Eq.2.1} and \eqref{con:Eq.2.2} restricted to the optimal domain $\Omega\left(\varphi\right)$.
     
     \noindent \textbf{Proof.}
     Shown in \ref{sec:D}.
     \label{lem:2.2}
\end{lemma}  
The following theorem establishes the existence and uniqueness of a minimum for the optimization problem \eqref{con:Eq.3}, based on Lemma \ref{lem:2.2}.

\begin{theorem}[\textbf{Existence and Uniqueness of a Minimum for the Optimization Problem}]
The optimization problem \eqref{con:Eq.3} admits a unique minimum.\\
\noindent \textbf{Proof.}
     The single objective functional $\mathcal{J}\left(\Omega,\psi_{D}\left(\Omega\right),\psi_{R}\left(\Omega\right)\right)$ is lower semicontinuous when the boundaries $\Sigma$ and $\Gamma$ are $C^{2}$.
     A continuous functional on a compact set attains its extrema.
     Therefore, by the compactness of $\mathcal{M}$ and the lower semicontinuity of $\mathcal{J}$, there exists a subsequence $\left(\Omega_{n_{k}},\psi_{D_{n_{k}}},\psi_{R_{n_{k}}}\right)$ from Lemma \ref{lem:2.2} such that
     \begin{equation}
          \lim_{k\rightarrow\infty} \mathcal{J}\left(\Omega_{n_{k}},\psi_{D_{n_{k}}},\psi_{R_{n_{k}}}\right)=\inf\left\{J\left(\Omega,\psi_{D}\left(\Omega\right),\psi_{R}\left(\Omega\right)\right):\left(\Omega,\psi_{D}\left(\Omega\right),\psi_{R}\left(\Omega\right)\right)\in\mathcal{M}\right\}.
          \label{con:Eq.41}
     \end{equation}
     By Lemma \ref{lem:2.2}, the sequence $\left(\Omega_{n_{k}},\psi_{D_{n_{k}}},\psi_{R_{n_{k}}}\right)$ converges to $\left(\Omega,\psi_{D},\psi_{R}\right)$,
     where $\psi_{D},\psi_{R}$ are defined on the optimal domain $\Omega$.
     $\hfill\blacksquare$
\label{thm:2}
\end{theorem} 
We therefore prove the well-posedness of the forward problem for PFSS extrapolation and the existence of a unique minimum for the associated inverse problem.
This work provides a mathematically rigorous argument for shape optimization based on PFSS extrapolation in coronal magnetic field modeling, 
and the proof can extend directly to nonspherical source surface geometries.

\section{Observations} \label{sec:3}
In this section, we describe the observations used to formulate the objective functional $\mathcal{J}$ and evaluate the optimization results for estimating $R_{ss}$. 
For a given $R_{ss}$, we calculate the PFSS extrapolation using a synoptic photospheric magnetogram as the inner boundary condition and the source surface at $r=R_{ss}$ as the outer boundary. 
We then use the radial solar wind velocity from PSP in situ measurements to trace measurements back to source surface footpoints along Parker spiral trajectories. 
PFSS extrapolation predicts the radial magnetic field at these footpoints, 
which are then scaled to the spacecraft position and compared with the observed radial magnetic field. 
This comparison defines the objective functional for estimating $R_{ss}$.
We also use ACE in situ measurements near 1 AU only as an independent dataset to validate $R_{ss}$ and solar cycle trend inferred from the optimization results based on PSP in situ measurements. 
We therefore describe below the magnetogram, PSP, and ACE observations used in this study.
\subsection{Synoptic Magnetograms} \label{subsec:3.1}

In this study, we use the zero point corrected Carrington rotation synoptic magnetograms from the Global Oscillation Network Group (GONG).
These magnetograms provide a homogeneous magnetic field boundary input for all PSP encounters considered here, 
have a grid resolution of $360\times180$ \cite{doi:10.1126/science.272.5266.1284}, 
and are updated daily, 
making them suitable for the repeated PFSS extrapolations required by the present inverse problem.

There are other products, 
such as synoptic magnetograms from the Helioseismic and Magnetic Imager (HMI) onboard the Solar Dynamics Observatory (SDO) \cite{pesnell_solar_2012,scherrer_helioseismic_2012}, 
where it could also be used for PFSS extrapolation.
For each candidate value of $R_{ss}$ in the optimization, 
PFSS extrapolation needs to be compared with in situ measurements.
Using HMI would either require retaining a higher effective spatial resolution, 
which increases the computational cost of repeated extrapolations, 
or applying additional rebinning or smoothing, 
which introduces another methodological choice beyond the scope of the present study as we specifically restrict our investigation to examination of in situ measurements and $R_{ss}$.
A comparison between optimizations driven by GONG and HMI magnetograms is therefore left for future work.

Table S1 in the Supporting Information illustrates the GONG synoptic magnetograms used for Carrington Rotations 2210-2282, 
covering PSP Encounters 1-19 and spanning the transition from solar minimum to the ascending phase of solar cycle 25.
During this period, PSP approached progressively closer to the Sun, 
with its perihelion distance decreasing to $11.4R_s$ by Encounter 19, 
thereby providing increasingly near Sun in situ constraints for the optimization of $R_{ss}$.

\subsection{PSP in situ Measurements} \label{subsec:3.2}

\begin{figure}[ht!]
      \includegraphics[width=35pc]{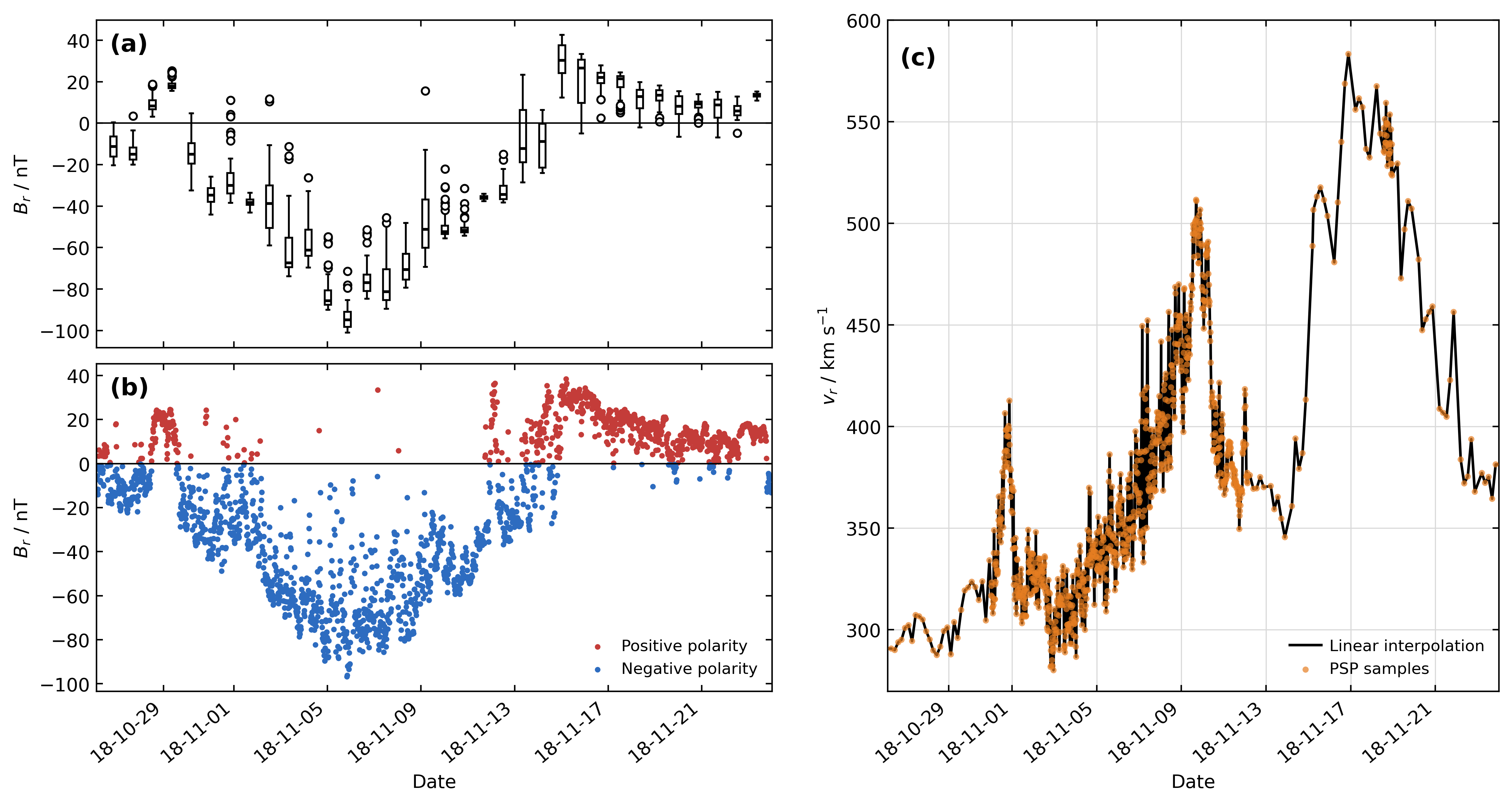}
      \caption{(a) PSP/FIELDS radial magnetic field $B_r$ measurements in hourly time bins for CR2210, 
      shown as boxplots used in the IQR outlier detection, with outliers marked by open circles. 
      (b) PSP/FIELDS $B_r$ after the IQR cleaning and 20 minutes averaging, 
      with red and blue denoting positive and negative polarity, respectively. 
      (c) PSP/SWEAP radial solar wind velocity $v_r$ used for Parker spiral backmapping. 
      The orange points show the PSP samples, 
      and the black curve shows the first order linear interpolation used to fill missing intervals. 
      The time axis spans the full interval of CR2210, whose perihelion occurred on 2018-11-05.}
      \label{fig:Fields}
\end{figure}

For each PSP encounter, the radial magnetic field and radial solar wind velocity are extracted from PSP in situ measurements during the Carrington rotation that contains the perihelion. 
This rotation-based selection provides a consistent temporal window for all encounters, 
emphasizes the near perihelion measurements most relevant to the optimization, 
and avoids extending the comparison into intervals with data gaps away from perihelion.
The FIELDS \cite{bale_fields_2016} and SWEAP \cite{kasper_solar_2016,Livi_2022} instruments provide the radial magnetic field and solar wind velocity measurements, respectively, as shown in Figure \ref{fig:Fields}.
These measurements are used to constrain the ballistic backmapping and the optimization of $R_{ss}$.  
Prior to the optimization, both measurements are averaged to a common temporal cadence. 
Outliers are removed from the magnetic field measurements, 
and data gaps in the solar wind velocity series are filled by linear interpolation to provide the continuous velocity input required for backmapping.

For the FIELDS magnetic field measurements, 
we divide the time series of each encounter into nonoverlapping 1-hour windows and apply the interquartile range (IQR) criterion to identify outliers \cite{tukey1977eda}. 
Within each window, the first and third quartiles, denoted by $Q_1$ and $Q_3$, 
are computed, and the interquartile range is defined as \(\mathrm{IQR}=Q_3-Q_1\). 
Measurements falling below \(Q_1-1.5\,\mathrm{IQR}\) or above \(Q_3+1.5\,\mathrm{IQR}\) are flagged as outliers and excluded from the subsequent analysis. 
Figure \ref{fig:Fields}(a) illustrates this boxplot-based outlier identification for 1-hour windows in CR2210, 
corresponding to PSP Encounter 1. 
After outlier removal, the remaining magnetic field measurements are averaged to a 20-minute cadence to reduce short timescale noise.
Figure~\ref{fig:Fields}(b) shows the resulting magnetic field time series after outlier removal and temporal averaging.

Although PSP provides dense coverage during each encounter, 
the radial solar wind velocity measured by SPAN-i contains intermittent data gaps. 
We first resample the velocity measurements to a 1-minute cadence and retain unavailable samples as NaN values. 
The resulting series is averaged into 20-minute bins, so that short gaps are effectively bridged when neighbouring valid measurements fall within the same bin, 
while longer intervals with no valid samples remain missing. 
These remaining gaps are filled by first order linear interpolation, 
producing a continuous velocity input required for Parker spiral backmapping. 
Figure~\ref{fig:Fields}(c) shows the measured and interpolated radial velocity profiles for CR2210, 
corresponding to PSP Encounter 1.

In principle, missing intervals in the SPAN-i radial velocity series could be reconstructed using physics-based or data-driven solar wind models, 
including the WSA model \cite{arge_improved_2003}, 
artificial neural network approaches \cite{https://doi.org/10.1029/2018SW001955,yang_modeling_2019}, 
and convolutional neural network approaches \cite{https://doi.org/10.1029/2023SW003561}. 
These methods are designed primarily to estimate large scale or global solar wind velocity distributions. 
In contrast, the present analysis requires only a continuous radial velocity series along the PSP trajectory, 
which is used as an input for Parker spiral backmapping from the spacecraft position to the source surface.
\ref{sec:B} presents the processed radial magnetic field and solar wind velocity data from PSP Encounters 1-19.

\subsection{ACE in situ Measurements} \label{subsec:3.3}

We use in situ measurements from the ACE spacecraft \cite{stone_advanced_1998} for independent cross-validation near 1 AU.
We obtain the ACE in situ measurements from the OMNI database and download the ACE Level 2 products AC-H2-MFI and AC-H2-SWI, 
which provide hourly measurements of the magnetic field and bulk velocity from the Magnetometer (MAG) and the Solar Wind Electron Proton Alpha Monitor (SWEPAM), respectively.
Using the encounter windows shown in Table S1, 
we extract 19 ACE intervals, each corresponding to the Carrington rotation that contains a PSP perihelion. 
We also apply the same processing procedure to the ACE measurements as to the PSP dataset to ensure consistency between the two in situ datasets.
Since ACE remains near the L1 point, it provides nearly continuous solar wind sampling over the selected Carrington rotations in a solar co-rotating coordinates. 
In the present study, ACE data serves as an independent cross-validation dataset for the optimization results inferred from PSP.

\section{Optimization Methodology} \label{sec:4}
This section constructs the optimization framework based on PSP in situ measurements that is used to estimate $R_{ss}$ for PFSS extrapolation. 
We first define the objective functional as the mean squared error (MSE) between PSP in situ radial magnetic field measurements and the corresponding PFSS extrapolation predictions.
For PSP in situ measurements, the spacecraft position is ballistically mapped back to footpoint on the source surface. 
The PFSS radial field predictions are then sampled at these footpoints and scaled radially to the PSP heliocentric distance, 
providing the model prediction used in the MSE.
We then describe the optimization algorithm to minimize the objective functional, 
validate the implementation with an analytical test problem, 
and define the additional metrics used to evaluate the optimization results.
\subsection{PFSS Extrapolation} \label{subsec:prior simulations}

The classical PFSS extrapolation solves Laplace equation, 
constrained by an input photospheric magnetogram and a zero potential spherical source surface. 
The spherical $R_{ss}$ serves as the only tunable parameter in this formulation. 
The PFSS extrapolation is widely employed in space weather modeling due to its computational efficiency, spatial resolution, and physically feasible approximation of the coronal magnetic structure \cite{Badman_2020}.
It is commonly solved by spherical harmonic expansion or the finite difference method (FDM) \cite{Toth_2011, Caplan_2021}. 
The spherical harmonic expansion offers the advantage of an explicit analytical solution, 
enabling direct evaluation of the magnetic potential at arbitrary spatial points without interpolation. 
It also introduces a free parameter of the maximum Legendre polynomial degree $L_{\rm max}$. 
Higher $L_{\rm max}$ mainly affects the radial magnetic field from the PFSS extrapolation at smaller heliocentric distances \cite{Toth_2011}.
FDM requires solutions on a grid followed by spatial interpolation. 
 
In this work, the inverse problem requires repeated PFSS extrapolations for different candidate values of $R_{ss}$ in the optimization.
Using FDM would therefore require repeated grid generation and interpolation in different footpoints during the optimization.
We therefore use the spherical harmonic expansion and fix $L_{\rm max}=22$ throughout the optimization.
This choice ensures that changes in the PFSS extrapolation reflect only the variation of $R_{ss}$ rather than harmonic expansion parameters $L_{\rm max}$.
The complete derivation of the PFSS extrapolation is detailed in \ref{sec:A}.

\subsection{Objective Functional} \label{subsec:Objective Functional}

For each candidate value of $R_{ss}$, 
we use radial solar wind velocity from PSP in situ measurements to trace the spacecraft position $\mathcal{O}$ to their footpoints $\mathcal{Q}=\mathcal{S}\left(\mathcal{O}\right)$ along Parker spiral lines. 
We construct this mapping $\mathcal{S}$ in Heliographic Carrington (HG) coordinates, 
and give the corresponding formulae in \ref{sec:B}. 
We denote the radial magnetic field $B_{r}^{\rm PFSS,k}$ as the PFSS extrapolation at the source surface footpoint for the same encounter index $k$.
Under the assumption used in the PFSS extrapolation that the magnetic field is radial on the source surface, 
the radial magnetic field predictions at the PSP position are obtained by scaling $R_{ss}$ with heliospheric distance $r_{\mathcal{O}}$ as,
\begin{equation}\label{magnetic equal}
     r_{\mathcal{O}}^{2}B_{r}^{\rm k}\left(\mathcal{O}\right)=R_{ss}^{2}B_{r}^{\rm PFSS,k}\left(\mathcal{S}\left(\mathcal{O}\right)\right)\quad \text{for $\rm k=1,2...,19$},
\end{equation}
where $B_{r}^{\rm k}$ are the corresponding radial magnetic field predictions from the PFSS extrapolations at the spacecraft position. 
We then calculate the MSE between the PSP in situ measurements $B_{r}^{\rm PSP, k}$ and $B_{r}^{\rm k}$ to construct the objective functional for the optimization algorithm.

To make the objective functional dimensionless for removing encounter dependent differences in mean level and amplitude, 
we first apply Z-score normalization to both $B_{r}^{\rm PSP, k}$ and $B_{r}^{\rm k}$ for each encounter $k$ as follows.
\begin{equation}\label{normalization}
       \tilde{B}_{r}^{\rm PSP,k}=\frac{B_{r}^{\rm PSP,k}-\mu^{\rm k}}{\sigma^{\rm k}},\quad
       \tilde{B}_{r}^{\rm k}=\frac{B_{r}^{\rm k}-\mu^{\rm k}}{\sigma^{\rm k}}
       \quad \text{for $\rm k=1,2,...,19$},
\end{equation}
where $\tilde{B}_{r}^{\rm PSP,k}$ and $\tilde{B}_{r}^{\rm k}$ denote the normalized observed and modeled radial magnetic field, 
and $\mu^{\rm k}$ and $\sigma^{\rm k}$ are the mean and standard deviation of the $B_{r}^{\rm PSP, k}$. 
This normalization subtracts the encounter-specific mean and rescales the magnetic field time series by its characteristic amplitude before the MSE is evaluated. 
The objective functional measures the agreement in the normalized temporal variation rather than being dominated by differences in the absolute magnetic field amplitude among encounters. 
It also brings the objective values and their gradients to similar numerical ranges for different encounters, 
which improves the conditioning of the optimization problem and allows fixed numerical settings, 
such as the step size and stopping tolerance, to be applied more consistently across encounters.

For encounter $k$, the continuous objective functional is defined as
\begin{equation}\label{RMSE}
     \tilde{\mathcal{J}}^{\rm k}=\frac{1}{2}\int_{I_k}\left|\tilde{B}_{r}^{\rm k}\left(\mathcal{O}_{t}\right)-\tilde{B}_{r}^{\rm PSP,k}\left(\mathcal{O}_{t}\right)\right|^{2}dt\quad\text{for $\rm k=1,2,...19$},
\end{equation}
where $I_k$ denotes the observation interval for the kth PSP encounter.
For the discrete time series used in the optimization, the objective functional is defined as
\begin{equation}\label{disRMSE}
     \mathcal{J}^{\rm k}=\frac{1}{2N_{k}}\sum_{i=1}^{N_{k}}\left|\tilde{B}_{r}^{\rm k}\left(\mathcal{O}_{t_i}\right)-\tilde{B}_{r}^{\rm PSP,k}\left(\mathcal{O}_{t_i}\right)\right|^{2}\quad\text{for $\rm k=1,2,...19$},
\end{equation}
where $N_k$ is the number of valid time samples in the kth encounter.

\subsection{Optimization Strategy} \label{sec:Optimization Strategy}
\subsubsection{Optimization Algorithm} \label{subsec:Optimization Algorithm}
\begin{algorithm}
\caption{Gradient Based Optimization of $R_{ss}$ with PSP in situ Measurements}
\label{alg:rss_optimization}
\begin{algorithmic}[1]
\Require 
    $B_{r}^{\rm PSP,k}$ (radial magnetic field from PSP's FIELDS) \\
    $v_{r}^{\rm PSP,k}$ (radial velocity from PSP's SWEAP)\\
    $B_{\gamma}^{\rm k}$ (Carrington heliospheric magnetograms from GONG)\\
    $R_{ss}^{0}$ (initial $R_{ss}$) \\
    $N_k$ (number of valid time samples in encounter $k$)\\
    $\mu^{\rm k}$ (mean value of $B_{r}^{\rm PSP,k}$ for each encounter)\\
    $\sigma^{\rm k}$ (standard deviations of $B_{r}^{\rm PSP,k}$ for each encounter)\\
    $\Delta h$ (finite difference increment used to approximate $\nabla \mathcal{J}$ by centered differences)\\
    $\eta$ (gradient descent step size) \\
    $N_{\rm max}$ (maximum number of iterations) \\
    $\epsilon$ (convergence criterion)
\Ensure $R_{ss}^{\rm opt}$ (optimized $R_{ss}$)

\State $p \gets 0$ \Comment{set the iteration index to zero}
\State $ R_{ss}^{p}$ $\gets$ $R_{ss}^{0}=1.700R_{s}$ \Comment{set the initial $R_{ss}$}
\Repeat
    \State $\mathcal{J} \gets 0$ \Comment{initialize the objective functional $\mathcal{J}$}
    \State $\nabla \mathcal{J} \gets 0$ \Comment{initialize the gradient $\nabla \mathcal{J}$}
    \State $B_{r}^{\rm PFSS,k} \gets \mathrm{PFSS}\!\left(B_{\gamma}^{\rm k}, R_{ss}^{p}\right)$ \Comment{compute the PFSS extrapolation for the current trial radius}
    
    \For{$t = 1$ \textbf{to} $N_k$} \Comment{loop over the valid samples in encounter $k$}
        \State $\mathcal{O}_{t} \gets \text{PSP orbit}$ \Comment{read the PSP position in HG coordinates}
        
        \State $\mathcal{Q}_{t} \gets \mathcal{S}\left(\mathcal{O}_{t}\right)$ from $v_{r}^{\rm PSP,k}\left(t\right)$ and $R_{ss}^{p}$ \Comment{Parker spiral tracing to the source surface}
        
        \State $B_{r}^{\rm PFSS,k}(\mathcal{Q}_{t}) \gets B_{r}^{\rm PFSS,k}, \mathcal{Q}_{t}$ \Comment{take $B_{r}^{\rm PFSS,k}$ at the traced $\mathcal{Q}_{t}$}
        
        \State $B_{r}^{\rm k}\left(\mathcal{O}_t\right)=B_{r}^{\rm PFSS,k}(\mathcal{Q}_{t})\cdot \left(\frac{R_{ss}^p}{r_{\mathcal{O}_t}}\right)^2$ \Comment{propagate the radial field to the spacecraft distance}
        
        \State $\tilde{B}_{r}^{\rm PSP,k}=\frac{B_{r}^{\rm PSP,k}-\mu^{\rm k}}{\sigma^{\rm k}},\tilde{B}_{r}^{\rm k}=\frac{B_{r}^{\rm k}-\mu^{\rm k}}{\sigma^{\rm k}}$ \Comment{apply Z score normalization}
        
        \State $\delta_{t} \gets \tilde{B}_{r}^{\rm PSP,k}\left(\mathcal{O}_{t}\right) - \tilde{B}_{r}^{\rm k}\left(\mathcal{O}_{t}\right)$ \Comment{evaluate the residual at sample $t$}
        
        \State $\mathcal{J} \gets \mathcal{J} + \delta_{t}^2$ \Comment{accumulate the squared data misfit}
    \EndFor
    
    \State $\mathcal{J} \gets \frac{\mathcal{J}}{2N_k}$ \Comment{evaluate the discrete objective functional}
    
    \State $R_{ss}^+ \gets R_{ss}^p + \Delta h$ \Comment{$\Delta h$ is the finite difference increment}
    \State $R_{ss}^- \gets R_{ss}^p - \Delta h$
    \State $\nabla \mathcal{J} \gets \frac{\mathcal{J}(R_{ss}^+) - \mathcal{J}(R_{ss}^-)}{2\Delta h}$ \Comment{centered finite difference approximation}
    
    \State $R_{ss}^{p+1} \gets R_{ss}^p - \eta \nabla \mathcal{J}$\Comment{update $R_{ss}$}
    
    \State $p \gets p + 1$
\Until{$|\nabla \mathcal{J}| < \epsilon$ \textbf{or} $p \geq N_{\rm max}$}

\State \Return $R_{ss}^p$ \Comment{return the optimized $R_{ss}$}
\end{algorithmic}
\end{algorithm}

The continuity of \(\mathcal{J}^{\rm k}\) with respect to \(C^{2}\) perturbations of the boundary provides the foundation for treating the determination of \(R_{ss}\) as a PDE-constrained inverse problem. 
Several numerical strategies could in principle be used for this class of problems, 
including classical gradient-based iterative methods \cite{Nurbekyan2022EfficientNG} 
and physics informed neural network that jointly optimize the PDE solution and unknown parameters 
\cite{RAISSI2019686,CUOMO202234,song_optimal_2024}. 
In the present study, the inverse problem involves only a single scalar parameter \(R_{ss}\). 
We therefore adopt a direct gradient-based iterative scheme, 
which is sufficient for the low dimensional optimization considered here and avoids the additional complexity of a neural network for PDE solver.
We solve the inverse problem about \(R_{ss}\) with the iterative procedure summarized in Algorithm \ref{alg:rss_optimization}. 
The algorithm minimizes the discrete objective functional $\mathcal{J}^{\rm k}$ in Eq.\ref{disRMSE} and returns the corresponding preferred $R_{ss}$. 

For a given \(R_{ss}^{p}\), 
the algorithm first calculates $B_{r}^{\rm PFSS,k}$ once, 
traces PSP position $\mathcal{O}_{t}$ to corresponding source surface footpoints $\mathcal{Q}_{t}$, 
samples the PFSS extrapolation $B_{r}^{\rm PFSS,k}$ at $\mathcal{Q}_{t}$,
radially scales $B_{r}^{\rm PFSS,k}\left(\mathcal{Q}_{t}\right)$ to $B_{r}^{\rm k}\left(\mathcal{O}_{t}\right)$ at the spacecraft position,
dimensionlessly normalizes $B_{r}^{\rm k}$ and $B_{r}^{\rm PSP,k}$,
and updates $\mathcal{J}^{\rm k}$. 
We then approximate the gradient with respect to \(R_{ss}\) by a centered finite difference with increment \(\Delta h\) and update the radius by
\[
R_{ss}^{p+1}=R_{ss}^{p}-\eta \nabla \mathcal{J}(R_{ss}^{p}).
\]
The iteration stops when \(|\nabla \mathcal{J}|<\epsilon\), 
which indicates that the update has approached a stationary point of the objective functional, 
or when the iteration count reaches the prescribed upper bound \(N_{\rm max}\).

To test the optimization algorithm, we construct a synthetic PFSS problem using an analytic potential field. 
A reference \(R_{ss}\) is prescribed, and radial magnetic field samples are generated from the analytic solution along the PSP Encounter 1 trajectory. 
The optimization is then applied to these synthetic data to determine whether the algorithm can recover the prescribed reference value of \(R_{ss}\). 
The algorithmic parameters, including the step size \(\eta\), tolerance \(\epsilon\), and maximum iteration number \(N_{\rm max}\), are adjusted in this test to assess convergence and robustness.

\subsubsection{Validation} \label{subsec:Analytical Validation}
\begin{figure}[ht!]
     \includegraphics[width=39pc]{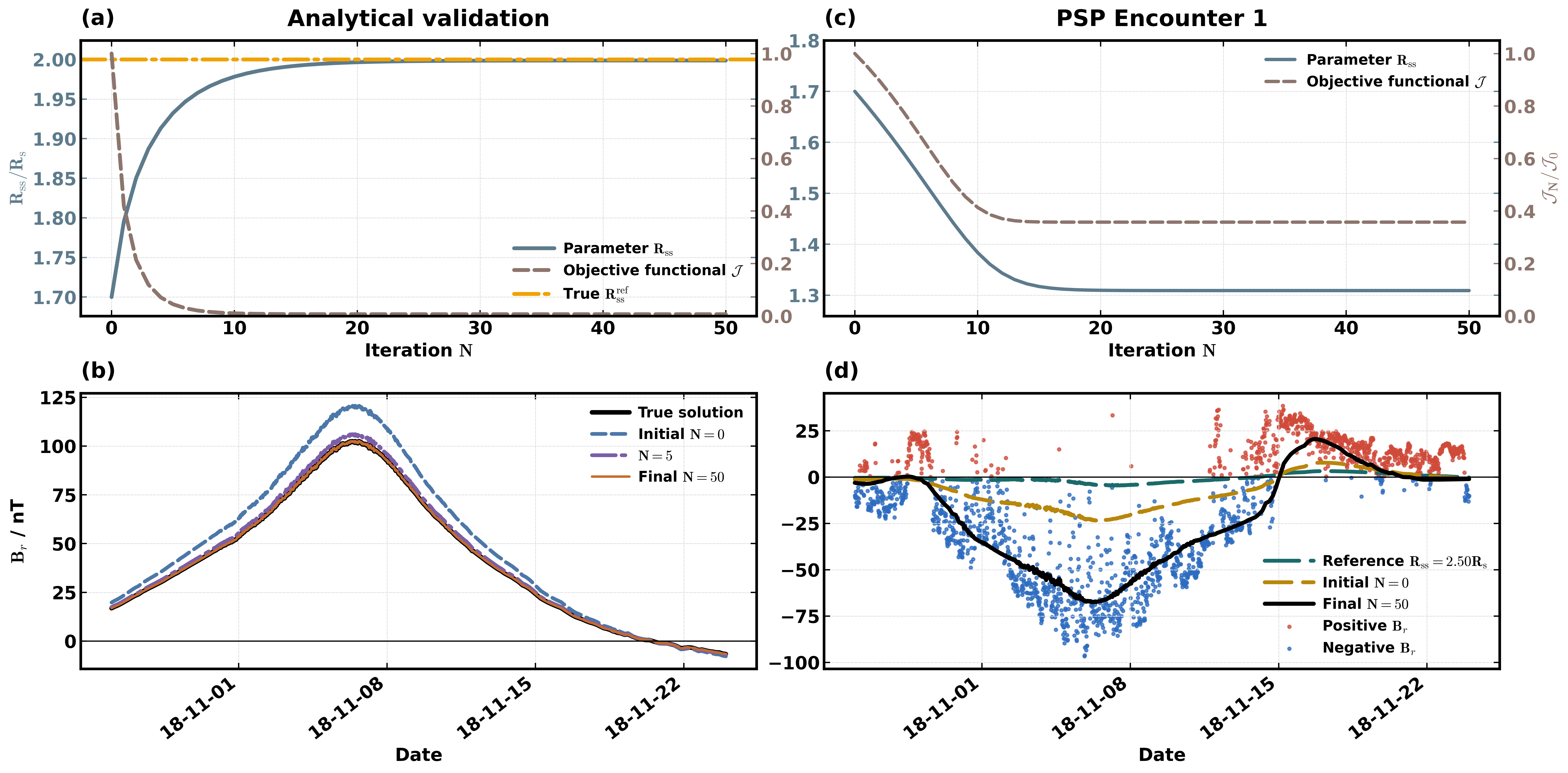}
     \centering
     \caption{(a) Iteration process for the analytical validation. 
     The blue gray curve shows the evolution of $R_{ss}$, 
     the brown dashed curve shows the normalized objective functional, 
     and the orange horizontal line marks the prescribed reference value $R_{ss}^{\rm ref}=2.0R_{s}$. 
     (b) Analytical reconstruction of $B_{r}$ at selected iterations. 
     The black solid curve denotes the true solution, 
     the blue dashed curve denotes the initial state $\rm N=0$, 
     the purple dash dot curve denotes the intermediate state $\rm N=5$, 
     and the orange solid curve denotes the final state $\rm N=50$. 
     (c) Iteration process for PSP Encounter 1. 
     The blue gray curve shows the evolution of $R_{ss}$, 
     and the brown dashed curve shows the normalized objective functional. 
     (d) Comparison between the $B_{r}^{\rm PSP,1}$ and $B_{r}^{\rm 1}$ for $2.5R_{s}$ and optimized solutions. 
     Red and blue points denote positive and negative observed $B_{r}^{\rm PSP,1}$, respectively; 
     the dark gray dashed curve denotes the reference case $R_{ss}=2.5R_{s}$, 
     the blue dashed curve denotes the initial state $\rm N=0$, 
     and the black solid curve denotes the final state $\rm N=50$.}
     \label{fig.Opt}
\end{figure}

Here we demonstrate and validate Algorithm \ref{alg:rss_optimization} both with an analytical test problem, 
and as applied to PSP Encounter 1 as an example. 
The results of these optimizations are shown in Figure \ref{fig.Opt}.
We first validate the numerical implementation with a test based on an analytic potential for the PFSS extrapolation \cite{Caplan_2021,10.3389/fspas.2022.1055969},
\begin{equation}
     \psi_{1}\left(r,\theta,\phi\right)=r\left[\left(\frac{1}{R_{ss}}\right)^{3}-\left(\frac{1}{r}\right)^{3}\right]\left(\cos\theta+\sin\theta \cos\phi\right),
     \label{con:Eq.43}
\end{equation}
where $r\in\left[1,R_{ss}\right]$, $\theta\in\left[0,\pi\right]$, $\phi\in[0,2 \pi)$.
We express \(R_{ss}\) in units of the solar radius \(R_s\), 
so that it is treated as a dimensionless optimization variable. 
The analytic potential \(\psi_1\) satisfies the Laplace equation with mixed boundary conditions, 
and the corresponding magnetic field is defined by \(\mathbf{B}=-\nabla\psi_1\). 
Its spherical components are given by
\begin{equation}
     \begin{cases}
          B_{r}\left(r,\theta,\phi\right)=\left[\left(R_{ss}\right)^{-3}+2r^{-3}\right]\left(\cos\theta+\sin\theta \cos\phi\right),\\
          B_{\theta}\left(r,\theta,\phi\right)=\left[r^{-3}-\left(R_{ss}\right)^{-3}\right]\left(\sin\theta-\cos\theta \cos\phi\right),\\
          B_{\phi}\left(r,\theta,\phi\right)=\left[r^{-3}-\left(R_{ss}\right)^{-3}\right]\sin\phi .
     \end{cases}
     \label{con:Eq.44}
\end{equation}

We use this analytic solution to construct a synthetic inverse problem for \(R_{ss}\). 
A reference value \(R_{ss}^{\rm ref}\) is prescribed, and synthetic radial magnetic field samples are generated from \(\psi_1\) along the PSP Encounter 1 trajectory. 
To mimic observational uncertainty, we add 20\% Gaussian noise to these synthetic samples. 
The same objective functional as in Algorithm~\ref{alg:rss_optimization} is then used to recover the prescribed value \(R_{ss}^{\rm ref}\).

Figure \ref{fig.Opt}(a) shows the convergence process of the optimization, 
while Figure\ref{fig.Opt}(b) compares the reconstructed radial magnetic field profiles obtained during the iterations with the synthetic reference data. 
For the case shown in Figure \ref{fig.Opt}(a), we set the initial guess to \(R_{ss}^{0}=1.700\), 
use the step size \(\eta=0.03\), and set the maximum number of iterations to \(N_{\rm max}=100\). 
The optimization converges to \(R_{ss}^{\rm ref}\) within 20 iterations. 
This analytic test therefore verifies that the numerical implementation can recover a prescribed \(R_{ss}\) from synthetic observations.

\begin{table}[ht!]
      \caption{Analytical Validation Cases for the Optimization Algorithm\label{tab:Testing}}
      \centering
      \begin{tabular}{lcccccc cccccc}
      \hline
      $l$ & $m$ & \(R_{ss}^{\rm ref}\) & Recovered Value & $g_{l}^{m}$ &$h_{l}^{m}$ &Iteration Cap\\
      \hline
      $1$ & $0$ & $2.000R_{s}$ & $2.003R_{s}$ & $1.288$ & $0.000$ & $150$\\ 
      $1$ & $1$ & $2.000R_{s}$ & $2.000R_{s}$ & $0.148$ & $-0.141$ & $150$ \\ 
      \hline
      $2$ & $0$ & $2.000R_{s}$ & $2.000R_{s}$ & $0.192$ & $0.000$ & $150$\\ 
      $2$ & $1$ & $2.000R_{s}$ & $2.001R_{s}$ & $-0.089$ & $0.075$ & $150$\\ 
      $2$ & $2$ & $2.000R_{s}$ & $1.998R_{s}$ & $-0.320$ & $-0.158$ & $150$ \\
      \hline
      $5$ & $0$ & $2.000R_{s}$ & $2.000R_{s}$ & $0.968$ & $0.000$ & $150$\\ 
      $5$ & $3$ & $2.000R_{s}$ & $2.000R_{s}$ & $0.500$ & $-0.157$ & $150$\\ 
      $5$ & $5$ & $2.000R_{s}$ & $2.000R_{s}$ & $0.427$ & $0.544$ & $150$\\
      \hline
      $10$ & $0$ & $2.000R_{s}$ & $1.999R_{s}$ & $-0.062$ & $0.000$ & $500$\\ 
      $10$ & $5$ & $2.000R_{s}$ & $2.001R_{s}$ & $-0.121$ & $-0.031$ & $500$\\ 
      $10$ & $10$ & $2.000R_{s}$ & $2.000R_{s}$ & $-0.463$ & $-0.143$ & $500$\\
      \hline
      \end{tabular}
      \begin{tablenotes} 
       \item This table lists analytical validation cases for selected spherical harmonic modes. 
      \end{tablenotes} 
\end{table}

We further validate the optimization algorithm using a family of analytic PFSS test fields constructed from individual spherical harmonic modes \cite{stansby2022testproblemspotentialfield}. 
For each degree \(l\) and order \(m\), with \(l\geq 1\) and \(0\leq m\leq l\), 
we define
\begin{equation}
     {\rm Test}
     =
     \left\{
     \mathbf{B}_{l}^{m}
     \ \bigg|\
     \mathbf{B}_{l}^{m}
     =
     \left(
     \left(B_{r}\right)_{l}^{m},
     \left(B_{\theta}\right)_{l}^{m},
     \left(B_{\phi}\right)_{l}^{m}
     \right)
     \right\},
     \label{con:Eq.45}
\end{equation}
where the spherical components are given by
\begin{equation}
     \left(B_{r}\right)_{l}^{m}\left(r,\theta,\phi\right)
     =
     c_{l}\left(r\right)P_{l}^{m}\left(\cos\theta\right)
     \left(g_{l}^{m}\cos m\phi+h_{l}^{m}\sin m\phi\right),
     \label{con:Eq.46}
\end{equation}
\begin{equation}
     \left(B_{\theta}\right)_{l}^{m}\left(r,\theta,\phi\right)
     =
     -d_{l}\left(r\right)
     \frac{\partial P_{l}^{m}\left(\cos\theta\right)}{\partial \theta}
     \left(g_{l}^{m}\cos m\phi+h_{l}^{m}\sin m\phi\right),
     \label{con:Eq.47}
\end{equation}
and
\begin{equation}
     \left(B_{\phi}\right)_{l}^{m}\left(r,\theta,\phi\right)
     =
     d_{l}\left(r\right)
     \frac{mP_{l}^{m}\left(\cos\theta\right)}{\sin \theta}
     \left(g_{l}^{m}\sin m\phi-h_{l}^{m}\cos m\phi\right).
     \label{con:Eq.48}
\end{equation}
Here \(P_l^m\) denotes the associated Legendre function, 
\(g_l^m\) and \(h_l^m\) are the mode coefficients, 
and the radial factors are
\begin{equation}
     c_{l}\left(r\right)
     =
     \left(\frac{R_{s}}{r}\right)^{l+2}
     \frac{
     \left(l+1\right)\left(R_{ss}\right)^{2l+1}+lr^{2l+1}
     }{
     \left(l+1\right)\left(R_{ss}\right)^{2l+1}+l\left(R_{s}\right)^{2l+1}
     },
     \label{con:Eq.49}
\end{equation}
\begin{equation}
     d_{l}\left(r\right)
     =
     \left(\frac{R_{s}}{r}\right)^{l+2}
     \frac{
     \left(R_{ss}\right)^{2l+1}-r^{2l+1}
     }{
     \left(l+1\right)\left(R_{ss}\right)^{2l+1}+l\left(R_{s}\right)^{2l+1}
     }.
     \label{con:Eq.50}
\end{equation}

Using this analytic test family, we examine individual modes with \(l=1,2,5,\) and \(10\), 
together with selected admissible orders \(m\). 
For each mode, synthetic radial magnetic field samples are generated with a prescribed reference value \(R_{ss}^{\rm ref}=2.0R_s\), 
and the optimization algorithm is then applied to recover \(R_{ss}\) from these data. 
As shown in Table \ref{tab:Testing}, 
the prescribed value \(R_{ss}^{\rm ref}=2.0R_s\) is recovered within the expected numerical tolerance for all listed cases. 
For the higher degree modes, we allow larger maximum iteration numbers, \(N_{\rm max}=150\) or \(500\), 
to ensure convergence under the chosen validation settings.

These tests verify that the optimization algorithm can recover a prescribed \(R_{ss}\) from analytic PFSS data across multiple spherical harmonic modes. 
They also clarify that, in this validation setting, the inferred change in open field behavior is controlled by the \(R_{ss}\) itself, 
while \(L_{\mathrm{max}}\) is kept fixed and does not serve as an additional optimization parameter.

We next apply the validated optimization algorithm to PSP Encounter 1 and use the numerical settings in Algorithm~\ref{alg:rss_optimization}: 
\(R_{ss}^{0}=1.70R_s\), \(\eta=0.5\), \(\epsilon=10^{-5}\), and \(N_{\rm max}=100\). 
Figures~\ref{fig.Opt}(c) and \ref{fig.Opt}(d) show the optimization result for PSP Encounter 1. 
Figure~\ref{fig.Opt}(c) presents the convergence process of \(R_{ss}\) during the iteration. 
Figure~\ref{fig.Opt}(d) compares $B_{r}^{\rm PSP,1}$ with two PFSS predictions: 
the optimized solution and the fixed reference solution obtained with the commonly used value 
\(R_{ss}=2.5R_s\). 
The optimization converges to a preferred value of \(R_{ss}=1.31R_s\) for Encounter 1. 
Compared with the fixed \(2.5R_s\) reference case, 
the optimized solution reduces the mismatch with $B_{r}^{\rm PSP,1}$. 
This inferred value is consistent with \(R_{ss}\) of approximately \(1.3R_s\) reported for PSP Encounter 1 by \citeA{Badman_2020}.

\subsection{Additional Evaluation Metrics} \label{subsec:Additional Evaluation Metrics}

We use two additional metrics to interpret the optimal \(R_{ss}\) and to identify the dominant factor controlling the optimization results obtained from Algorithm~\ref{alg:rss_optimization}. 
These metrics are the magnetic field scaling factor \(\alpha^{\rm k}\) and the polarity prediction accuracy \(P_a^{\rm k}\). 
The magnetic field scaling factor is defined as
\begin{equation}\label{scaling factor}
       \alpha^{\rm k} 
       =
       \max_{\mathcal O}
       \left|
       \frac{B_{r}^{\rm PSP,k}\left(\mathcal{O}\right)}
       {B_{r}^{\rm k}\left(\mathcal{O}\right)}
       \right|,
       \quad k=1,2,\ldots,19 ,
\end{equation}
where the maximum is taken over the valid samples in the \(k\)th encounter. 
This metric quantifies the amplitude correction required to bring \(B_{r}^{\rm k}\) to \(B_{r}^{\rm PSP,k}\) amplitude scale. 
For PFSS extrapolation, \(\alpha^{\rm k}\) is commonly larger than 1, 
reflecting the tendency of the model to underestimate the radial magnetic field strength. 
This tendency is closely related to the open flux problem, in which coronal magnetic field models underestimate the amount of open magnetic flux in the heliosphere \cite{Linker_2017}. 
Reducing \(R_{ss}\) generally increases the modeled open flux, 
thereby increasing \(B_{r}^{\rm k}\) amplitude. 
The scaling factor \(\alpha^{\rm k}\) therefore serves as an evaluation metric for interpreting the optimal \(R_{ss}\), 
if the optimization tends to favor \(R_{ss}\) that reduce the amplitude mismatch between \(B_r^{\rm k}\) and \(B_r^{\rm PSP,k}\). 

The polarity prediction accuracy \(P_a^{\rm k}\) is then used as another evaluation metric to assess whether the optimized solution also improves the agreement in magnetic polarity.
While \(\alpha^{\rm k}\) measures the amplitude discrepancy as a regression-driven metric between the PFSS prediction and the PSP measurements, 
\(P_a^{\rm k}\) depends only on the sign of \(B_r\) as a classification-driven metric. 
It therefore evaluates whether the model reproduces the magnetic sector structure and polarity reversals 
encountered along the spacecraft trajectory. 
We define
\begin{equation}\label{polarity}
     P_{a}^{\rm k}
     =
     \frac{1}{N_{k}}
     \sum_{i=1}^{N_{k}}
     \delta
     \left(
     \mathrm{sign}\left(B_{r}^{\rm k}\left(\mathcal{O}_{t_i}\right)\right),
     \mathrm{sign}\left(B_{r}^{\rm PSP,k}\left(\mathcal{O}_{t_i}\right)\right)
     \right),
     \quad k=1,2,\ldots,19 ,
\end{equation}
where \(N_k\) is the number of valid samples in the \(k\)th encounter, 
and \(\delta(a,b)=1\) if \(a=b\) and \(\delta(a,b)=0\) otherwise. 
Both \(\alpha^{\rm k}\) and \(P_a^{\rm k}\) are computed after the optimization. 
They are then used in the multiobjective analysis to determine whether the optimal \(R_{ss}\) are primarily driven by amplitude agreement or by polarity agreement.

\subsection{Pareto Optimal Algorithm}\label{subsec:Pareto Optimal Algorithm}

The single objective functional in Eq.~\ref{disRMSE} defines the optimization problem solved in Algorithm~\ref{alg:rss_optimization}. 
This choice is appropriate for inferring \(R_{ss}\) from PSP in situ measurements, 
but it does not exhaust the possible evaluations of PFSS extrapolations performance. 
In previous studies, PFSS extrapolations have been evaluated or constrained using several observational quantities, 
including the IMF strength, open magnetic flux, IMF polarity, 
coronal hole areas observed in AIA 193~\AA\ images, and magnetic topology inferred from TSE white light observations \cite{lee_coronal_2011,https://doi.org/10.1002/2013JA019464,Badman_2020,Habbal_2021,Badman_2022,Benavitz_2024}. 
Because these quantities measure different aspects of the heliospheric and coronal magnetic field, 
they may prefer different \(R_{ss}\). 
A value of \(R_{ss}\) that improves the amplitude agreement may not simultaneously maximize the polarity prediction accuracy.
Accordingly, after the single objective optimization has been completed,
we perform a multiobjective analysis to quantify the tradeoff between the evaluation metrics and to assess whether the optimized solutions are mainly driven by amplitude or polarity.

A candidate solution is Pareto optimal if one objective cannot be improved without worsening at least one other objective. 
The set of all such nondominated solutions is called the Pareto optimal set, and its image in the objective space is the Pareto frontier \cite{zbMATH01349589,zbMATH01614566}. 
In the present study, we use Pareto analysis as a posterior evaluation to examine the tradeoff between the amplitude and the polarity agreement. 
This analysis helps determine whether the optimal \(R_{ss}\) obtained from Algorithm~\ref{alg:rss_optimization} are primarily associated with improved consistency.

For each PSP encounter, we consider the parameter interval \(\mathcal{X}=[1.2R_s,3.0R_s]\) and define two objectives. 
The first objective is the single objective functional \(\mathcal{J}^{\rm k}\) defined in Eq.~\ref{disRMSE}. 
The second objective is the polarity prediction error, \(1-P_a^{\rm k}\), 
where \(P_a^{\rm k}\) is defined in Eq.~\ref{polarity}. 
Thus, the multiobjective problem is written as
\begin{equation}
     \min_{R_{ss}\in\mathcal{X}}
     \left(
     \mathcal{J}^{\rm k}(R_{ss}),
     1-P_a^{\rm k}(R_{ss})
     \right),
     \label{eq:pareto_objectives}
\end{equation}
where both components are minimized.

The scaling factor \(\alpha^{\rm k}\) is not included as an objective in the Pareto construction. 
Instead, it is retained as a dominant metric for interpreting the resulting Pareto solutions and the optimal \(R_{ss}\). 
The polarity objective is piecewise constant and does not satisfy the continuity assumptions used in Theorem~\ref{thm:2}. 
In practice, we sample \(R_{ss}\) over \(\mathcal{X}\) for PSP Encounters 1-19, 
evaluate the two objectives at each sampled value, and identify the nondominated points to obtain a discrete approximation of the Pareto optimal set and its frontier. 
We next define the dominance relation used in this construction.

\begin{definition}[Dominance Relation]
For two candidate solutions $R_{ss}^{(1)}, R_{ss}^{(2)} \in \mathcal{X}$, we say that $R_{ss}^{(1)}$ \emph{dominates} $R_{ss}^{(2)}$, denoted as $R_{ss}^{(1)} \prec R_{ss}^{(2)}$, if and only if
\begin{align}
\mathcal{J}^{\rm k}\left(R_{ss}^{(1)}\right) \leq \mathcal{J}^{\rm k}\left(R_{ss}^{(2)}\right), \quad
1-P_{a}^{\rm k}\left(R_{ss}^{(1)}\right) \leq 1-P_{a}^{\rm k}\left(R_{ss}^{(2)}\right), \quad \text{and} \\
\mathcal{J}^{\rm k}\left(R_{ss}^{(1)}\right) < \mathcal{J}^{\rm k}\left(R_{ss}^{(2)}\right) \quad \text{or} \quad
1-P_{a}^{\rm k}\left(R_{ss}^{(1)}\right) < 1-P_{a}^{\rm k}\left(R_{ss}^{(2)}\right) & \label{eq:dominance}
\end{align}
\end{definition}
\begin{definition}[Nondominated Set]
A set $\mathcal{ND} \subseteq \mathcal{X}$ is called a \emph{non dominated set} if
\[
\forall R_{ss}^{(1)}, R_{ss}^{(2)} \in \mathcal{ND}, \quad R_{ss}^{(1)} \nprec R_{ss}^{(2)} \text{ and } R_{ss}^{(2)} \nprec R_{ss}^{(1)}
\]
That is, no solution in $\mathcal{ND}$ dominates any other solution in $\mathcal{ND}$.
\end{definition}

\begin{definition}[Pareto Optimal Set]
The \emph{Pareto optimal set} $\mathcal{P}^*$ is the maximal non dominated subset of $\mathcal{X}$ such that
\[
\mathcal{P}^* = \left\{ R_{ss}^* \in \mathcal{X} \mid \nexists R_{ss} \in \mathcal{X} : R_{ss} \prec R_{ss}^* \right\}.
\]
Equivalently, $\mathcal{P}^*$ consists of all solutions that are not dominated by any other solution in $\mathcal{X}$.
\end{definition}

\begin{definition}[Pareto Frontier]
The \emph{Pareto frontier} $\mathcal{F}^*$ is the image of the Pareto optimal set in the objective space
\[
\mathcal{F}^* = \left\{ \left(\mathcal{J}^{\rm k}(R_{ss}^*), 1-P_{a}^{\rm k}(R_{ss}^*)\right) \in \mathbb{R}^2 \mid R_{ss}^* \in \mathcal{P}^* \right\}.
\]
\end{definition}

Several algorithms can construct the Pareto optimal set, 
including the layering method, the banker's method, 
the divide and conquer method, and the quicksort method. 
We use a quicksort based construction because, 
after we evaluate the objectives on a discretized parameter set, 
we can extract the nondominated solutions efficiently from the sorted list.
\begin{algorithm}
\caption{Quicksort Method for Multiobjective Optimization}
\label{alg:pareto}
\begin{algorithmic}[2]
\Require 
    Parameter space $\mathcal{X}=[R_{\min}, R_{\max}]$, number of evaluations $N$
\Ensure Pareto optimal set $\mathcal{P}^*$, Pareto frontier $\mathcal{F}^*$

\State Discretize parameter space so that $R_{ss}^{(i)} = R_{\min} + \frac{i-1}{N-1}(R_{\max} - R_{\min})$, $i=1,\dots,N$
\For{$i = 1$ \textbf{to} $N$}
    \State Calculate $\mathcal{J}^{\rm k,(i)} = \mathcal{J}^{\rm k}(R_{ss}^{(i)})$ using Equation (\ref{disRMSE})
    \State Calculate $E_{P}^{\rm k,(i)} = 1-P_{a}^{\rm k}(R_{ss}^{(i)})$ using Equation (\ref{polarity})
\EndFor
\State Sort indices by $\mathcal{J}^{\rm k}$, namely $\mathbf{I} = \text{argsort}(\{\mathcal{J}^{\rm k,(1)},\dots,\mathcal{J}^{\rm k,(N)}\})$
\State Initialize $\mathcal{P}^* = \emptyset$ and $\min E_{P}^{\rm k} = \infty$
\For{$t = 1$ \textbf{to} $N$}
    \State $i = I_t$
    \If{$E_{P}^{\rm k,(i)} < \min E_{P}^{\rm k}$}
        \State $\mathcal{P}^* \leftarrow \mathcal{P}^* \cup \{R_{ss}^{(i)}\}$
        \State $\min E_{P}^{\rm k} = E_{P}^{\rm k,(i)}$
    \EndIf
\EndFor
\State Construct $\mathcal{F}^* = \{(\mathcal{J}^{\rm k,(i)}, E_{P}^{\rm k,(i)}) : R_{ss}^{(i)} \in \mathcal{P}^*\}$
\Return $\mathcal{P}^*, \mathcal{F}^*$
\end{algorithmic}
\end{algorithm}

\subsection{Reference Simulations}\label{subsec:Reference Simulations}

Before applying the optimization, 
we first perform reference PFSS extrapolations using two fixed \(R_{ss}\) at \(2.0R_s\) and \(2.5R_s\) for PSP Encounters 1-19. 
These reference cases are evaluated using the same objective functional and evaluation metrics as those used for the optimized solutions. 
The corresponding results for the PSP and ACE datasets are summarized in Supporting Information Tables S2 and S3, respectively. 
Here the PSP comparisons provide the direct baseline for the optimization, 
whereas the ACE comparisons serve as an independent 1 AU validation of the inferred \(R_{ss}\). 
Together, these fixed \(R_{ss}\) simulations provide reference cases against which the improvement achieved by the optimization can be assessed.

\section{Results}\label{sec:result}

This section summarizes the optimization results for PSP Encounters 1-19. 
We begin by presenting the \(R_{ss}\) values inferred from Algorithm~\ref{alg:rss_optimization} 
and relating their evolution to the transition from solar minimum to the ascending phase of solar cycle 25. 
We then examine deviations from this overall solaractivity trend and evaluate the optimized solutions 
using additional metrics and Pareto analysis. 
As an independent validation, ACE observations from the corresponding Carrington rotations are used to assess 
that the results derived from PSP remain consistent with 1 AU measurements.

\subsection{Single Objective Optimization}\label{subsec:Rsspspace}

\begin{figure*}[p]
      \centering
      \includegraphics[width=\textwidth,height=0.90\textheight,keepaspectratio]{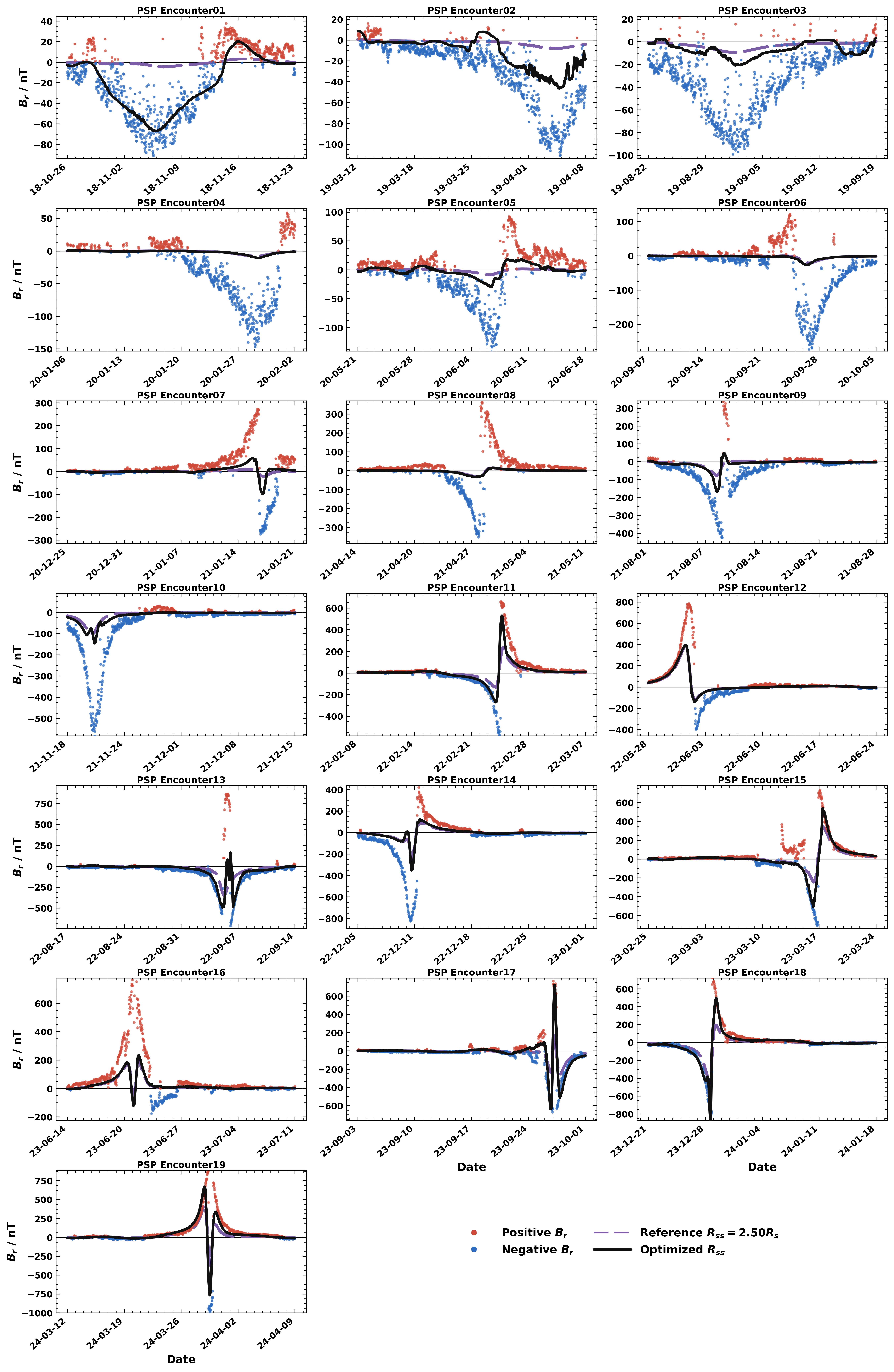}
      \caption{
      Comparison between $B_r^{\rm PSP,k}$ and $B_r^{\rm k}$ for Encounters 1-19. 
      Purple curves show the reference solution with $R_{ss}=2.50R_s$, 
      black curves show the optimized solution from Algorithm \ref{alg:rss_optimization}, 
      and red and blue points denote positive and negative $B_r^{\rm PSP,k}$, respectively.}
               \label{fig:Simulation}
          \end{figure*}

We apply Algorithm \ref{alg:rss_optimization} by minimizing the objective functional $\mathcal{J}$ in Eq.\ref{disRMSE}. 
Each run is initialized at $1.70R_{s}$ and terminated when the convergence criterion is met or when $N_{\rm max}$ is reached. 
Figure \ref{fig:Simulation} compares the optimized solutions for PSP Encounters 1-19 with the fixed reference case $R_{ss}=2.5R_{s}$. 

The optimized solutions generally show better agreement with \(B_r^{\rm PSP,k}\) than the fixed \(R_{ss}\) reference case. 
The inferred \(R_{ss}\) values also exhibit an overall increase from solar minimum toward the ascending phase of solar cycle 25. 
This trend can be understood from the amplitude agreement shown in Figure~\ref{fig:Simulation}. 
For the earlier encounters, especially those before Encounter 9, the fixed \(R_{ss}\) reference solution lies well below the upper envelope of $B_r^{\rm PSP,k}$. 
This indicates a stronger underestimation of the amplitude by the reference PFSS extrapolation during solar minimum and the early ascending phase. 
To reduce this amplitude discrepancy, the optimization favors a lower \(R_{ss}\), 
which increases the amount of open magnetic flux and strengthens $B_r^{\rm k}$ along the PSP trajectory. 
For later encounters, the reference solution is closer to the observed upper envelope, 
so a less required reduction of \(R_{ss}\) is required. 
The resulting optimized \(R_{ss}\) therefore increases toward the ascending phase.

\begin{figure}[ht!]
      \centering
      \includegraphics[width=35pc]{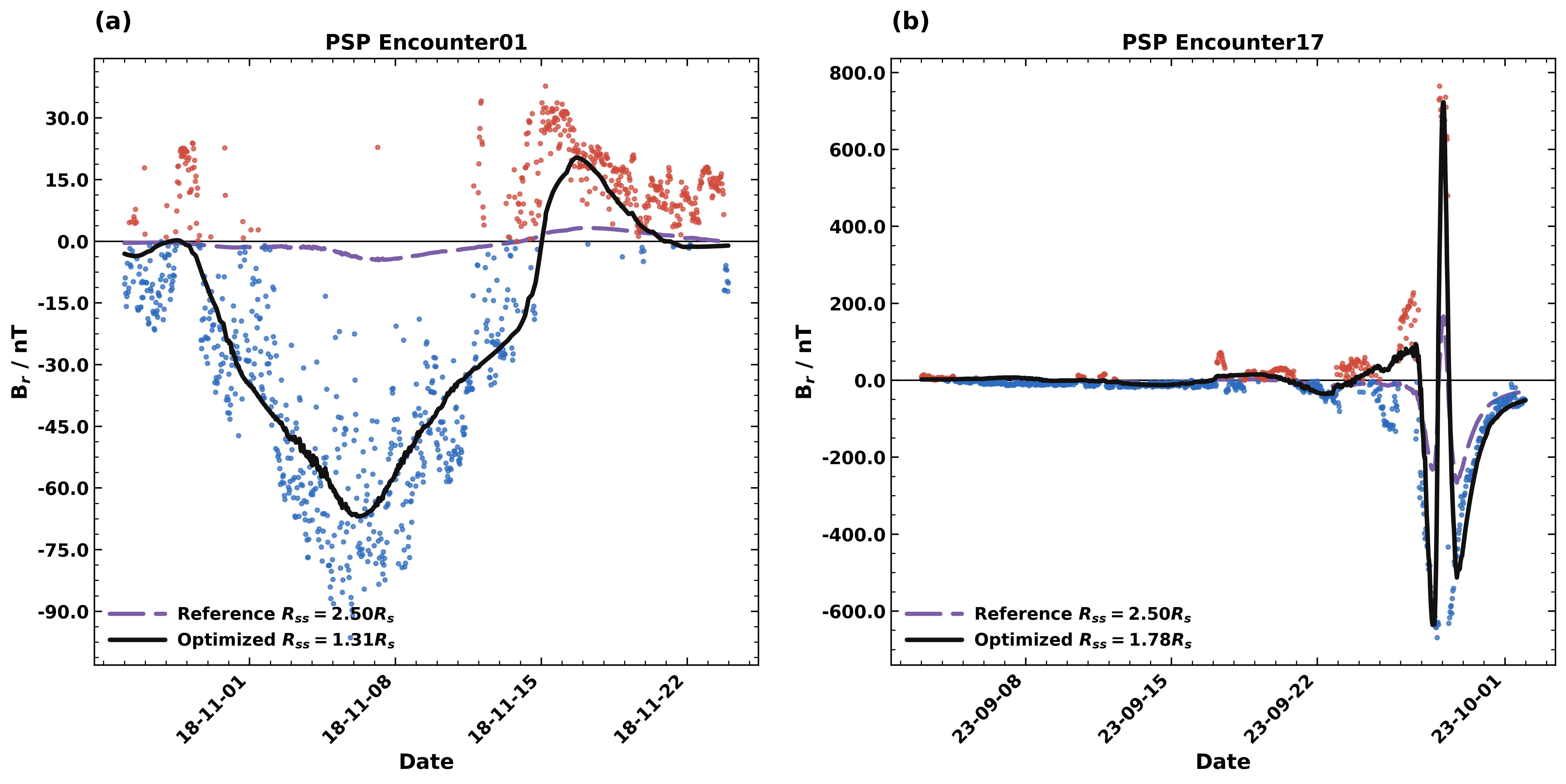}
      \caption{(a) Comparison between $B_r^{\rm PSP,k}$ and $B_r^{\rm k}$ for PSP Encounter 1 during solar minimum. 
      Red and blue points denote positive and negative $B_r^{\rm PSP,k}$, respectively. 
      The purple dashed curve denotes the reference $B_r^{\rm k}$ with $R_{ss}=2.5R_{s}$, 
      and the black solid curve denotes the optimized solution. 
      (b) Comparison for PSP Encounter 17 during the ascending phase.}
      \label{fig:ak}
\end{figure}

This behavior is illustrated in Figure~\ref{fig:ak}. 
Figure~\ref{fig:ak}(a) compares the reference and optimized PFSS predictions with $B_r^{\rm PSP,k}$ for Encounter 1, 
representative of solar minimum conditions, 
and Figure~\ref{fig:ak}(b) shows the corresponding comparison for Encounter 17 during the ascending phase. 
Because the reference PFSS field is substantially weaker than the observed field during Encounter 1, 
the optimization favors a lower \(R_{ss}\) than in Encounter 17, where the reference solution is already closer 
to the upper envelope of $B_r^{\rm PSP,k}$.

Not all early encounters follow this amplitude agreement behavior. 
Within the \(k<9\) group, Encounters 4, 6, and 8 show only limited visible improvement in field amplitude relative to the fixed \(R_{ss}\) reference solution. 
For these cases, the optimal \(R_{ss}\) appears to be less strongly controlled by the amplitude agreement and more strongly influenced by the polarity agreement as shown in Table S2. 
This indicates that the dominant factor selecting the optimal \(R_{ss}\) can vary between encounters, 
with amplitude agreement dominating most early encounters and polarity agreement becoming more important for Encounters 4, 6, and 8.

\begin{figure}[ht!]
      \centering
      \includegraphics[width=30pc]{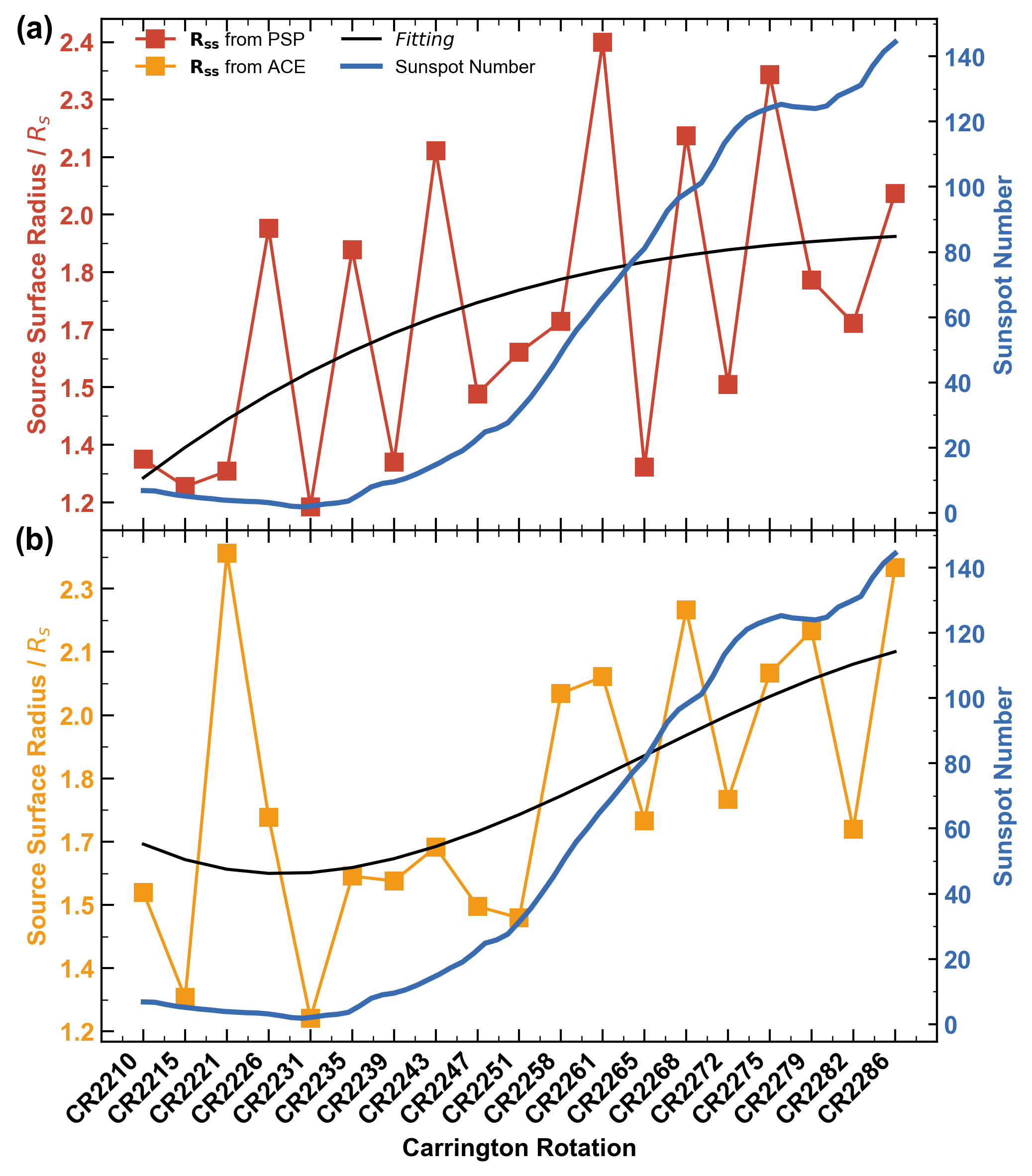}
      \caption{(a) The optimal $R_{ss}$ for PSP Encounters 1-19. 
      The red polyline with square markers shows the encounter by encounter optimization results, 
      the black curve shows the spline used only to visualize the trend, 
      and the blue curve shows the sunspot number over the same Carrington rotation range on the secondary vertical axis. 
      (b) The optimal $R_{ss}$ for ACE datasets. 
      The orange polyline with square markers shows the ACE derived optimal $R_{ss}$ values, 
      the black curve shows the corresponding spline trend, 
      and the blue curve again shows the sunspot number on the secondary vertical axis.}
      \label{fig:Rsspspace}
\end{figure}

Figure~\ref{fig:Rsspspace}(a) shows the optimal $R_{ss}$ obtained from PSP Encounters 1-19.
The red polyline represents the optimization results across the encounters,
and the black spline \cite{DIERCKX1975165} is shown as a guide to the overall trend.
The blue curve gives the contemporaneous sunspot number as a proxy for solar activity.
The $R_{ss}$ generally increase with solar activity,
with lower $R_{ss}$ during solar minimum and larger values during the ascending phase of solar cycle 25.
Figure~\ref{fig:Rsspspace}(b) presents the corresponding results obtained from the ACE datasets.
The $R_{ss}$ from the ACE datasets exhibit a similar temporal evolution, 
providing an independent 1 AU validation of the trend inferred from PSP.

\subsection{Local Departures from the Overall Trend}\label{subsec:polarity inverse}

\begin{figure}[ht!]
     \centering
     \includegraphics[width=38pc]{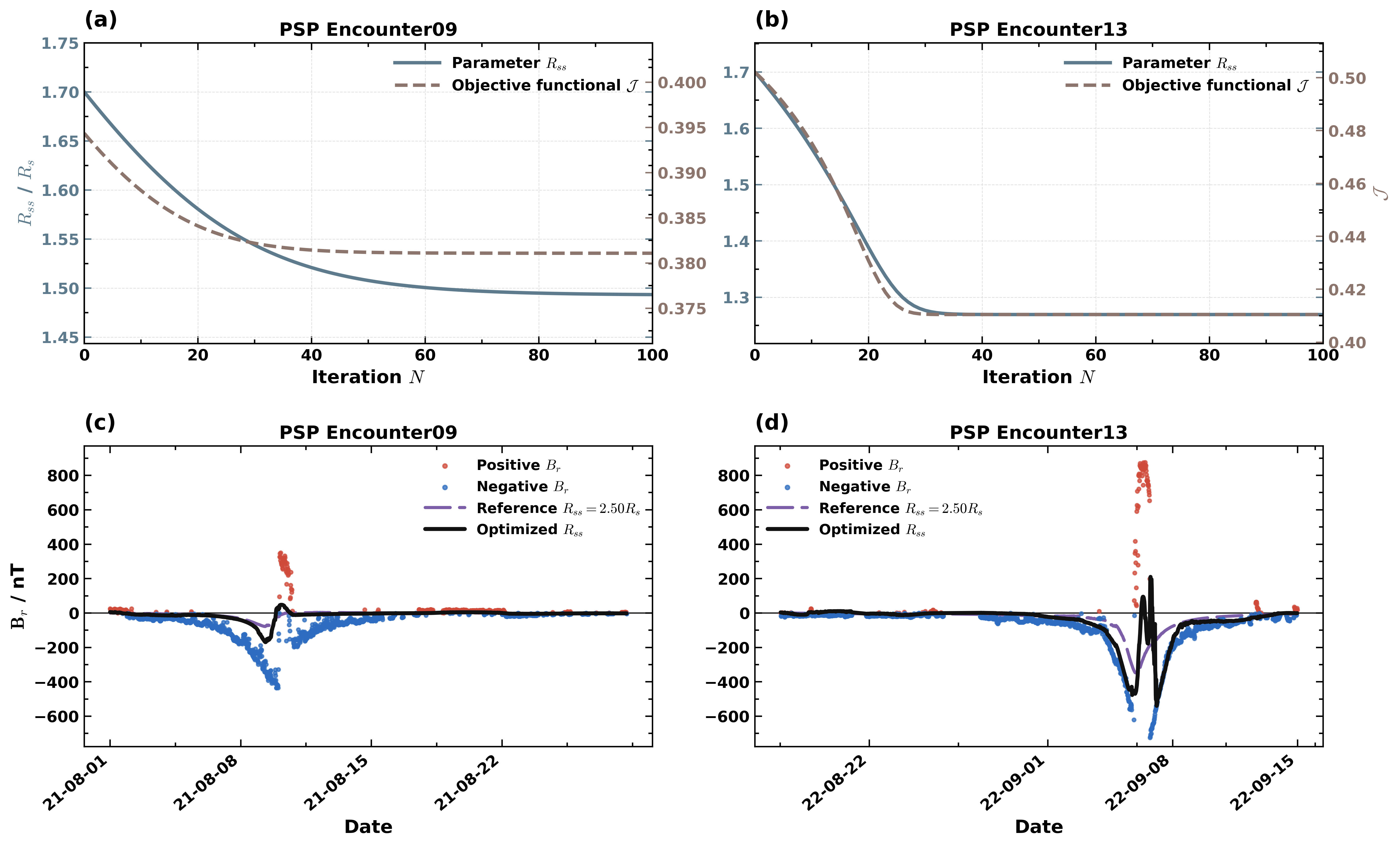}
     \caption{(a) Iteration process for PSP Encounter 9. 
     The blue gray curve shows the evolving parameter $R_{ss}$, 
     and the brown dashed curve shows the objective functional $\mathcal{J}$. 
     (b) Iteration process for PSP Encounter 13, 
     with the same curve definitions as in panel (a). 
     (c) Comparison between the $B_r^{\rm PSP,k}$ and $B_r^{\rm k}$ for Encounter 9. 
     Red and blue points denote positive and negative observed $B_{r}$, respectively. 
     The purple dashed curve denotes the reference $B_r^{\rm k}$  with $R_{ss}=2.50R_{s}$, 
     and the black solid curve denotes the optimized $B_r^{\rm k}$ . 
     (d) Comparison for Encounter 13, with the same curve definitions as in panel (a).}
     \label{fig.U}
\end{figure}

Although the overall trend shows a positive correlation between the optimal $R_{ss}$ and solar activity, 
this trend is not monotonic for every Carrington rotation. 
Figure \ref{fig.U} highlights two ascending phase examples, Encounters 9 and 13, 
for which the optimization selects lower $R_{ss}$ values than the surrounding encounters. 

Figure \ref{fig.U}(a) shows that the optimization for Encounter 9 decreases $R_{ss}$ gradually while the objective functional drops rapidly at early iterations and then approaches a plateau. 
Figure \ref{fig.U}(c) shows that the optimized solution with $R_{ss}=1.49R_{s}$ captures more polarity inversions than the reference solution.
The analysis of Encounter 9 (see Figure \ref{fig:openflux}) suggests that the local departures from the overall trend are driven by the polarity prediction accuracy.
This is because $B_r^{\rm PSP,k}$ for this encounter contains more complex polarity inversions than the other encounters which reference $R_{ss}=2.5R_{s}$ fails to capture, 
and the optimization selects lower $R_{ss}$ values to capture these inversions.
These conditions favor inserting more polarity inversions which drives a lower $R_{ss}$. 

Figure \ref{fig.U}(b) shows a similar decrease of $R_{ss}$ for Encounter 13, 
but Figure \ref{fig.U}(d) indicates a stronger local departure because the time series contains more complex reversals around 5 September 2022 and 6 September 2022. 
During this interval, 
PSP crossed an interplanetary coronal mass ejection (ICME) characterized by complex topological structures \cite{Romeo_2023b2}. 
This event drives the optimal $R_{ss}$ to $1.27R_{s}$. 
The application of optimization algorithm to events with CME compromises the meaningful estimation of the optimal $R_{ss}$. 
Under these circumstances,
spacecraft observations occur within a severely perturbed interplanetary environment where the magnetic field configuration breaks the Parker spiral approximation.

\subsection{Open Flux and Polarity Accuracy}\label{open flux}
\begin{figure}[ht!]
     \centering
     \includegraphics[width=38pc]{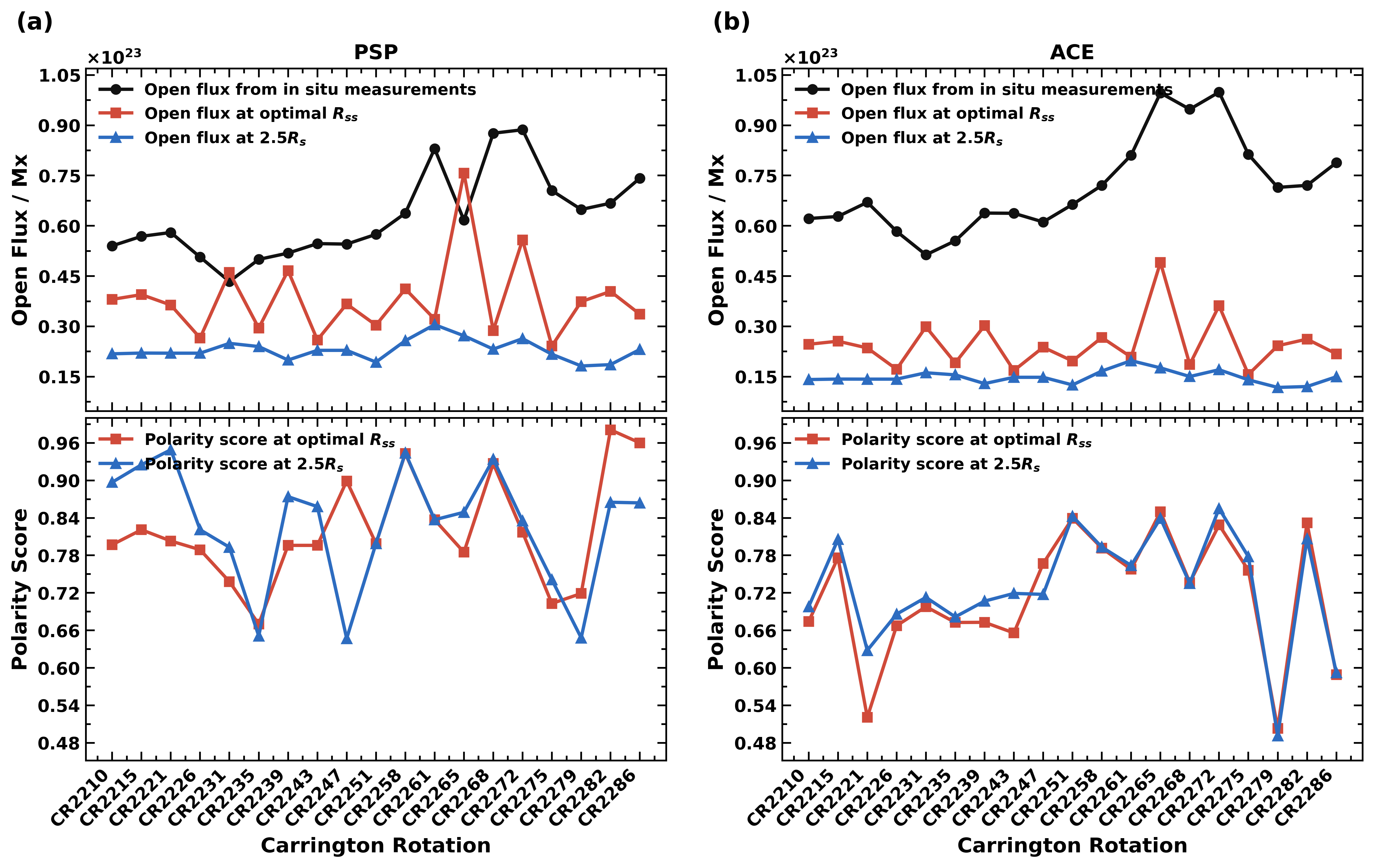}

     \caption{(a) Open flux and polarity prediction accuracy for the PSP based evaluation. 
     The upper plot compares the open flux inferred from PSP in situ measurements, 
     shown by the black curve, 
     with the optimized solution in red and the fixed reference case with $R_{ss}=2.5R_{s}$ in blue. 
     The lower plot gives the corresponding polarity prediction accuracy for the optimized and reference solutions. 
     (b) Open flux and polarity prediction accuracy for the ACE based evaluation. 
     The upper plot shows the ACE open flux comparison, 
     where the black curve denotes the in situ estimate and the red and blue curves denote the optimized and reference solutions. 
     The lower plot shows the associated polarity prediction accuracy.
     Circular markers denote the open flux inferred from in situ measurements, square markers denote the optimized solution, and triangular markers denote the reference solution.
     \label{fig:openflux}}
\end{figure}

We focus here on the two metrics that are not used in the optimization, 
the open flux \(\Phi_{\rm flux}^{\rm k}\), 
a meaningful quantity where the ratio of the predicted and measured value is equal to $\alpha^{\rm k}$, 
as defined in section \ref{subsec:Additional Evaluation Metrics}, 
and the polarity prediction accuracy \(P_{a}^{\rm k}\).
In the PFSS extrapolation, 
the global open flux is obtained by integrating the radial magnetic field over the source surface
\begin{equation}
     \Phi_{\rm flux}^{\rm k}\left(R_{ss}\right)=R_{ss}^{2}\int_{0}^{2\pi}d\phi\int_{0}^{\pi}\left|B_{r}^{\rm PFSS,k}\left(R_{ss},\theta,\phi\right)\right|\sin\theta\,d\theta.
     \label{con:Eq.51}
\end{equation}
The latitude independent nature of \(B_{r}^{\rm k}\) \cite{https://doi.org/10.1029/95GL02826,smith_open_2003} suggests that the open magnetic flux estimated from in situ measurements can be formulated as
\begin{equation}
     \Phi_{\rm H}^{\rm k}=\frac{4\pi}{N}\sum_{i=1}^{N} \left(r_{i}^{\rm PSP,k}\left(\mathcal{O}_i\right)\right)^{2}\left|B_{r}^{\rm PSP,k}\left(\mathcal{O}_i\right)\right|.
     \label{con:Eq.52}
\end{equation}

Figure \ref{fig:openflux}(a) shows that the open flux derived from PSP in situ measurements exceeds the open flux from the PFSS extrapolation for both the optimized solution and the fixed reference case $R_{ss}=2.5R_{s}$. 
The optimized solution shows improved agreement with the in situ open flux compared to the reference case,
indicating that the optimization process effectively reduces the open flux discrepancy.
The lower plot in Figure \ref{fig:openflux}(a) shows the polarity prediction accuracy $P_{a}^{\rm k}$. 
For encounters with $k<9$, $P_{a}^{\rm k}$ has a small reduction relative to the reference case, which indicates that the optimization process slightly reduces polarity prediction accuracy for improved open flux agreement.
That is, the optimization process for these encounters is more driven by the amplitude agreement than by the polarity agreement.
$P_{a}^{\rm k}$ is not the dominant metric in selecting the optimal $R_{ss}$ for these encounters.
During CR2247, or Encounter 9, the optimized solution improves polarity prediction accuracy which has been the dominant metric for optimization. 
For encounters with $k\geq 9$, $P_{a}^{\rm k}$ remains or improves accuracy.
Section \ref{subsec:Pareto} interprets this behavior as a change in the balance between amplitude agreement and polarity agreement.

Figure \ref{fig:openflux}(b) shows the corresponding ACE evaluation. 
It follows the same trend as the PSP results and shows higher polarity prediction accuracy during the ascending phase than during solar minimum. 
That is the lower sensitivity of $P_{a}^{\rm k}$ to $R_{ss}$ during the ascending phase.
Figure S1 illustrates the global distribution of the source surface magnetic field across varying levels of solar activity. 
During the ascending phase, the sector structure and the heliospheric current sheet tend to be more inclined relative to the ecliptic, 
so an ecliptic spacecraft crosses a sector boundary becomes more predictable and less sensitive to modest changes in $R_{ss}$. 
During solar minimum, the current sheet is flatter, and small changes in the extrapolated field or mapped trajectory can move the spacecraft from one side of the sector boundary to the other, 
which makes $P_{a}^{\rm k}$ more sensitive to $R_{ss}$ \cite{https://doi.org/10.1029/2023JA031359}. 
Polar field measurements remain poorly constrained from the ecliptic plane, 
which affects open flux estimates \cite{Yang_2024}. 
This limitation is most important during solar minimum, 
when long lived polar coronal holes dominate the global magnetic structure \cite{Wang_2009, Hahn_2010, karna_using_2014, pishkalo_polar_2019, ANDREEVA20231915, AYang_2024}. 
Magnetograph calibration \cite{Wang_2010_Mag} and time dependent contributions from coronal hole boundaries \cite{Arge_2024_APJ} may also affect the comparison.

\subsection{Pareto Optimal Set for Multiobjective Optimization}\label{subsec:Pareto}

\begin{figure}[ht!]
     \centering
     \includegraphics[width=38pc]{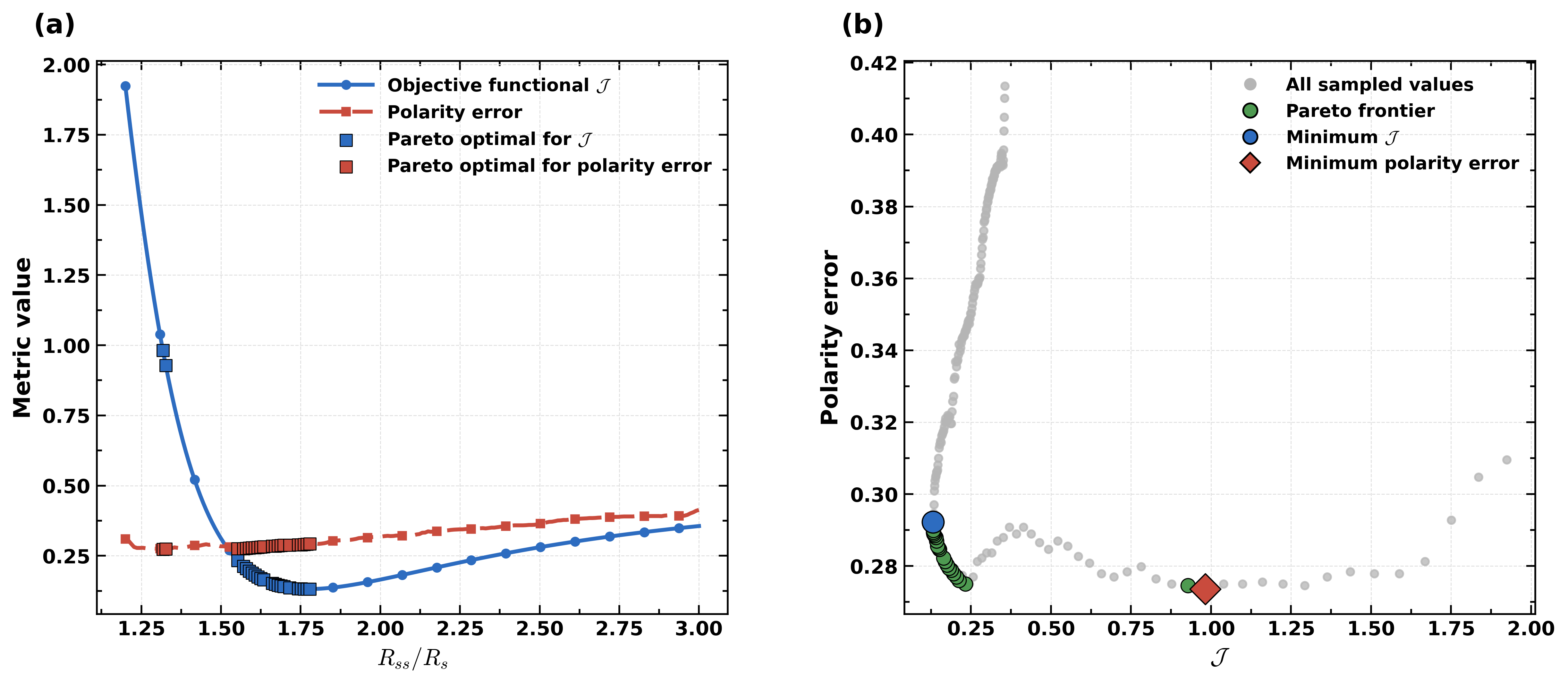}

     \caption{(a) Pareto analysis for PSP Encounter 17 in parameter space. 
     The blue curve shows the objective functional $\mathcal{J}^{\rm k}$ from Eq.\ref{disRMSE}, 
     and the red dashed curve shows the polarity prediction error $1-P_{a}^{\rm k}$ from Eq.\ref{polarity} as functions of $R_{ss}$. 
     The blue and red square markers denote the Pareto optimal subsets associated with these two metrics.
     (b) Pareto frontier for PSP Encounter 17 in the two objective space. 
     The gray points represent all sampled candidate values of $R_{ss}$, 
     the green circles denote the Pareto frontier, 
     the blue circle marks the global minimum of the objective functional $\mathcal{J}^{\rm k}$, 
     and the red diamond marks the global minimum of the polarity prediction error.\label{fig:C1Pareto}}
\end{figure}

We use the Pareto optimal set to evaluate which metric plays the dominant role in selecting the final value of \(R_{ss}\) and to clarify why the optimal \(R_{ss}\) varies across encounters.
The Pareto optimal set consists of the nondominated candidate values of \(R_{ss}\) in the parameter space,
whereas the Pareto frontier is the image of these candidates in the two-objective space \((\mathcal{J}^{\rm k},1-P_a^{\rm k})\).
A more dispersed Pareto frontier indicates a stronger conflict between the MSE objective and the polarity prediction error:
reducing the MSE requires sacrificing polarity agreement, or improving polarity agreement requires a larger MSE.
In this case, the optimization is mainly controlled by the amplitude agreement.
By contrast, a more clustered Pareto frontier indicates that the nondominated candidates occupy a narrower region of the objective space.
When the MSE values of these candidates become less separated, 
the polarity prediction error becomes more influential in determining which \(R_{ss}\) is preferred.

Figure~\ref{fig:C1Pareto} illustrates this analysis for Encounter 17.
Figure~\ref{fig:C1Pareto}(a) shows the variation of the objective functional and the polarity prediction error with \(R_{ss}\),
together with the corresponding Pareto optimal set in the parameter space.
Figure~\ref{fig:C1Pareto}(b) maps the same sampled candidate values into the two-objective space and shows the resulting Pareto frontier.
Because the MSE objective and the polarity prediction error do not attain a common global minimum,
the nondominated candidates form a Pareto optimal set rather than a single solution that simultaneously minimizes both objectives.
For Encounter 17, the Pareto frontier is relatively clustered, indicating that the admissible tradeoffs are concentrated within a narrow objective range.
In this case, the final choice of \(R_{ss}\) is more strongly constrained by polarity agreement than by amplitude agreement.

\begin{figure*}[p]
      \centering
      \includegraphics[width=\textwidth,height=0.90\textheight,keepaspectratio]{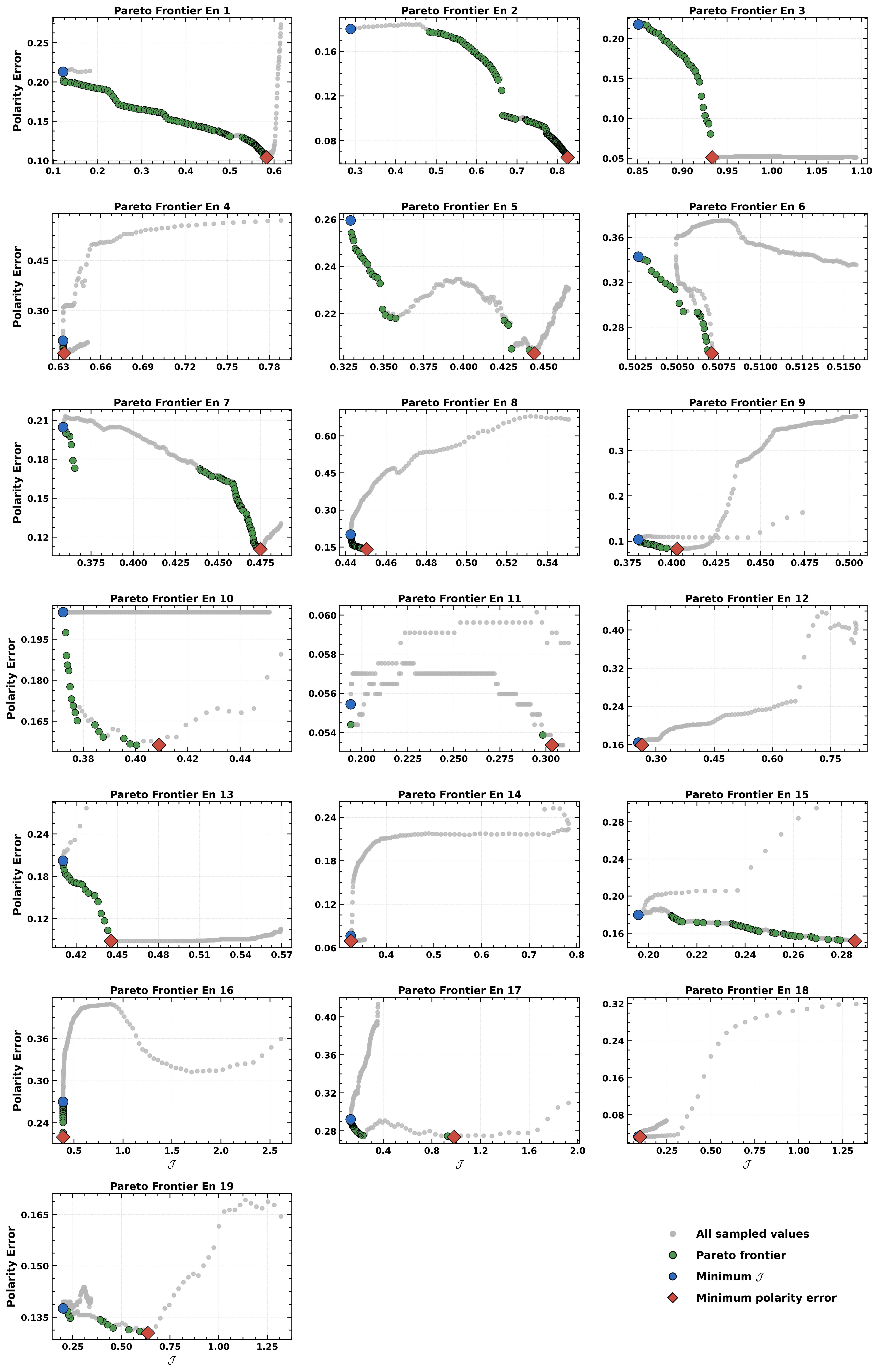}
      \caption{
      Pareto frontiers for PSP Encounters 1-19 calculated with Algorithm \ref{alg:pareto}. 
      In each panel, the gray points represent all sampled candidate values of $R_{ss}$ in the two objective space, 
      the green circles denote the Pareto frontier, 
      the blue circles mark the minima of the objective functional, 
      and the red diamonds mark the minima of the polarity prediction error.
      Across the encounter sequence, 
      the frontiers evolve from more dispersed configurations to more clustered ones.}
      \label{fig:Paretoset}
\end{figure*}

We then apply Algorithm~\ref{alg:pareto} to PSP Encounters 1-19 and show the resulting Pareto frontiers in Figure~\ref{fig:Paretoset}.
The gray points denote all sampled values of \(R_{ss}\) projected into the objective space \((\mathcal{J}^{\rm k},1-P_a^{\rm k})\),
and the green points mark the nondominated candidates that define the discrete Pareto frontier.
Each green point represents a candidate \(R_{ss}\) for which one objective cannot be improved without degrading the other.
The configuration of the frontier therefore reveals the degree of competition between amplitude agreement and polarity agreement.

For the earlier encounters, the Pareto frontiers are more dispersed, indicating a stronger conflict between the two objectives.
The polarity metric does not dominate the selection of \(R_{ss}\); 
instead, the optimization is mainly driven by amplitude agreement.
As solar activity increases, the Pareto frontiers become more clustered in the objective space.
This clustering shows that the range of nondominated tradeoffs becomes narrower and that polarity prediction error becomes increasingly important in selecting the preferred \(R_{ss}\).
The transition from dispersed to clustered Pareto frontiers therefore indicates a shift in the dominant metric of the optimization,
from amplitude agreement during solar minimum toward polarity agreement during the ascending phase of solar cycle 25.

\section{Conclusion} \label{sec:5}
This paper completes the mathematical theory proof and constructs an optimization algorithm for the optimal source surface problem, 
along with optimization validation and multiobjective optimization analysis.

Section \ref{sec:2} proves the well-posedness of the forward problem for PFSS extrapolation and reformulates the inverse problem for $R_{ss}$ as a free boundary problem. 
By combining the compactness of the admissible set with continuity of the objective functional, 
we show that the inverse problem admits at least one minimizer. 
The arguments can apply to nonspherical source surface geometries as well as to the spherical case used in the present PSP application.

Section \ref{sec:4} also develops and presents Algorithm \ref{alg:rss_optimization} to perform the optimization. 
We define the objective functional as a normalized MSE between $B_r^{\rm PSP,k}$ and $B_r^{k}$. 
The algorithm then minimizes this objective functional, 
and the analytical validation shows that a prescribed reference radius can be recovered from noisy test data. 
Additional evaluation metrics, including the magnetic field scaling factor and polarity prediction accuracy, 
are used to interpret the optimized solutions. 
The Pareto analysis extends this interpretation by quantifying the tradeoff between the objective functional and polarity prediction error.

Overall, the optimal \(R_{ss}\) inferred from PSP Encounters 1-19 increase from solar minimum to the ascending phase of solar cycle 25. 
The same trend is recovered from ACE measurements over the corresponding Carrington rotations. 
Relative to the fixed reference cases, the optimal \(R_{ss}\) reduce the open flux discrepancy 
and generally preserve or improve the magnetic field polarity agreement. 
Some encounters depart from the overall trend, 
especially when polarity reversals or CME disturbed magnetic structure make lower \(R_{ss}\) values more favorable under the adopted PFSS extrapolation and Parker spiral assumptions.

Finally, we perform a multiobjective analysis using the MSE objective \(\mathcal{J}\) and the polarity prediction error \(1-P_a\). 
Algorithm \ref{alg:pareto} is applied to construct the Pareto frontiers for PSP Encounters 1-19. 
These frontiers are used to evaluate the relative importance of amplitude agreement and polarity agreement in selecting the optimal \(R_{ss}\), 
and thereby to interpret the mechanisms underlying the variation of \(R_{ss}\) across the encounters.
This transition from dispersed to clustered Pareto frontiers indicates a shift in the dominant metric from amplitude agreement during solar minimum toward polarity agreement during the ascending phase. 
It therefore explains why the optimized \(R_{ss}\) increases as solar activity strengthens.

By comparisons with TSE white-light topology, relevant studies have represented that optimization according to TSE white light images yield a best $R_{ss}$ which instead decreases during solar maximum. \cite{altschuler_magnetic_1969,Benavitz_2024}. 
Taking these results together shows there is no globally optimal $R_{ss}$, 
but instead that optimization is with respect to the predefined metrics.
Our optimization enhances both polarity predictions and open flux estimations at various solar activity cycles and indicates that during the solar ascending phases, a higher $R_{ss}$ yields better predictions of polarity and open flux, whereas improved white-light morphology predictions require the opposite conditions.
This discrepancy is summarized by the open flux problem, 
where (1) physical model cannot simultaneously satisfy multiple independent observations and (2) estimates of open flux from coronal hole observations conflict with those from in-situ measurements \cite{Linker_2021,Asvestari_2024}.
Besides, we do not incorporate observations from the solar polar regions, which may impose constraints on the estimation of $R_{ss}$ during solar minimum conditions.
On the other hand, PFSS extrapolation demonstrates an underestimation of open magnetic flux predictions, 
which necessitates improvements in coronal model or magnetograms boundary condition \cite{Linker_2017}. 
This underestimation may be improved by incorporating photospheric magnetic field observations from solar polar regions. 
In fact, continuous in-situ measurements for both northern and southern polar magnetic fields in the inner corona region would improve the performance of our optimization algorithm.
Currently, the Solar Orbiter is scheduled to reach an orbital inclination of 33 degrees by July 2029, 
which will provide support for further redefinement of our optimization framework \cite{SolarOrbiter2022}.
Our future work will focus on two aspects: on the one hand, refining the theoretical framework of shape optimization for coronal magnetic field models; on the other hand, reformulating the coronal magnetic field model to address these limitations.

\section*{Conflict of Interest}
The authors declare no conflicts of interest relevant to this study.
\section*{Open Research}
The PFSS extrapolation and optimization algorithm code is available at \url{https://doi.org/10.5281/zenodo.17129425} \cite{wang_2025_17129425}.
PSP magnetic field data are available at \url{https://fields.ssl.berkeley.edu/} and solar wind velocity data are accessible at \url{http://sweap.cfa.harvard.edu/}. 
GONG magnetogram data can be obtained from the National Solar Observatory (NSO) at \url{https://nso.edu/data/nisp-data/}.
ACE data are available via the OMNIWeb \url{https://omniweb.gsfc.nasa.gov/}.
All calculations and visualizations in this paper are implemented by using Python 3.9.13, 
leveraging the numpy (v1.25.2) \cite{harris2020array}, scipy (v1.11.1) libraries \cite{2020SciPy-NMeth} for array operations and associated Legendre function computations
and astropy.io.fits (v5.3.1) for CDF/FITS format data \cite{2022ApJ...935..167A}, 
sunpy.map (v5.0.0) for Carrington magnetograms \cite{sunpy_community2020}
and matplotlib (v3.7.2) for visualizations, available under the Matplotlib license \cite{Hunter:2007}.

\begin{acknowledgments}
      The work is jointly supported by the NSFC and the National Key R\&D Program of China (42330210, 2022YFF0503800, 42004146 and 2021YFA0718600),the Specialized Research Fund for State Key Laboratories, the open project fund of State Key Laboratory of Lunar and Planetary Sciences (Macau University of Science and Technology) (Macau FDCT grant No. 002/2024/SKL), and the Climbing Program of NSSC (E5PD3003).
      Magnetic field and velocity measurements from the Parker Solar Probe can be accessed through FIELDS and SWEAP, respectively. 
      We obtain the solar wind data measured by the ACE spacecraft from the OMNI database.
      This work uses GONG data obtained by the NSO Integrated Synoptic Program, managed by the National Solar Observatory, which is operated by the Association of Universities for Research in Astronomy (AURA), Inc. under a cooperative agreement with the National Science Foundation and with contributions from the National Oceanic and Atmospheric Administration. The GONG network of instruments is hosted by the Big Bear Solar Observatory, High Altitude Observatory, Learmonth Solar Observatory, Udaipur Solar Observatory, Instituto de Astrofisica de Canarias, and Cerro Tololo Interamerican Observatory.
      The authors gratefully acknowledge these supports.
\end{acknowledgments}

\bibliography{agujournaltemplate}
\clearpage

\appendix
\setcounter{figure}{0}
\setcounter{table}{0}
\setcounter{equation}{0}
\renewcommand{\thefigure}{A\arabic{figure}}
\renewcommand{\thetable}{A\arabic{table}}
\renewcommand{\theequation}{A\arabic{equation}}
\section{Magnetic Field at Source Surface and PSP in situ Measurements}\label{sec:A}

Figure S1 in Supporting Information shows the solution of Eq.\ref{con:Eq.A.1} at the source surface. Table S1 lists the Carrington rotation magnetograms used for PSP Encounters 1-19. The vertical axis gives colatitude and the horizontal axis gives longitude. We follow the spherical harmonic implementation described in Xudong Sun's notes, available at \url{http://wso.stanford.edu/words/pfss.pdf}. Equations (10), (16), (17), and (18) in those notes contain errors. The magnetic potential series must start from the zeroth harmonic term, $L=0$, and the magnetic field series must start from the first harmonic term, $L=1$. The forward problem for PFSS extrapolation is
\begin{equation}
     \begin{cases}
          -\Delta \psi=0 & \text{in $\Omega$,} \\
          \partial_{\mathbf{n}} \psi=B_r\left(\gamma\right) & \text{on $\Gamma$,} \\
          \psi=0 & \text{on $\Sigma$.} \\
     \end{cases} 
     \label{con:Eq.A.1}
\end{equation}
This mixed boundary elliptic PDE can be solved by separation of variables.
For a spherical source surface, the solution of the PFSS extrapolation has the analytic expansion
\begin{equation}
     \psi\left(r,\theta,\phi\right)=R_{s}\sum_{l=0}^{\infty}\sum_{m=0}^{l}P_{l}^{m}\left(\cos\left(\theta\right)\right)\left(g_{l}^{m}\cos\left(m\phi\right)+h_{l}^{m}\sin\left(m\phi\right)\right)\left[c_{l}\left(\frac{R_{s}}{r}\right)^{l+1}-d_{l}\left(\frac{r}{R_{s}}\right)^{l}\right],
     \label{con:Eq.A.2}
\end{equation}
where $c_{l},d_{l}$ are given by
\begin{equation}
     c_{l}=\frac{\left(R_{ss}\right)^{2l+1}}{\left(l+1\right)\left(R_{ss}\right)^{2l+1}+l\left(R_{s}\right)^{2l+1}},\quad d_{l}=\frac{\left(R_{s}\right)^{2l+1}}{\left(l+1\right)\left(R_{ss}\right)^{2l+1}+l\left(R_{s}\right)^{2l+1}}.
     \label{con:Eq.A.3}
\end{equation}
The magnetic field is defined by $\mathbf{B}=-\nabla\psi$,
and the coefficients $g_{l}^{m},h_{l}^{m}$ are projected on the triangular basis functions
\begin{equation}
     B_{r}\left(\gamma\right)=\sum_{l=1}^{\infty}\sum_{m=0}^{l}P_{l}^{m}\left(\cos\left(\theta\right)\right)\left(g_{l}^{m}\cos\left(m\phi\right)+h_{l}^{m}\sin\left(m\phi\right)\right)\left[\left(l+1\right)c_{l}+ld_{l}\right],
     \label{con:Eq.A.4}
\end{equation} 

\begin{equation}
     \int_{\Gamma}B_{r}\left(\gamma\right)H_{l}^{m}\left(\gamma\right)d\Gamma=
     \begin{cases}
          2\pi& \text{$m=0$},\\
          \pi& \text{$m\neq 0$},
     \end{cases}
     \label{con:Eq.A.5}
\end{equation} 
where $H_{l}^{m}\left(\gamma\right)=P_{l}^{m}\left(\cos\left(\theta\right)\right)\cos m\phi$,
the associated Legendre functions $P_{l}^{m}\left(x\right)$ are normalized with the coefficients
\begin{equation}
     A_{l}^{m}=\sqrt{\frac{2l+1}{2}\frac{\left(l-m\right)!}{\left(l+m\right)!}},\quad \int_{-1}^{1}\left|A_{l}^{m}P_{l}^{m}\left(x\right)\right|^{2}dx=1.
     \label{con:Eq.A.6}
\end{equation}

We process the radial solar wind velocity and radial magnetic field from PSP in situ measurements with first order linear interpolation and the IQR method, respectively. 
Figure S2 in Supporting Information exhibits the processed results for each encounter, 
with the left vertical axis of each subplot quantifying the radial solar wind velocity ($\rm km/s$) and the right vertical axis representing the radial magnetic field magnitude ($nT$). 
This dual axis visualization enables simultaneous analysis of velocity and magnetic field correlations across distinct encounters.
The subplots reveals that the observed polarity inversions exhibit correlations with captured high velocity streams during Encounters 11, 13-17. 
This association suggests that the ascending phases of solar activity influences the iteration of $R_{ss}$ within optimization frameworks.


\setcounter{figure}{0}
\setcounter{table}{0}
\setcounter{equation}{0}
\renewcommand{\thefigure}{B\arabic{figure}}
\renewcommand{\thetable}{B\arabic{table}}
\renewcommand{\theequation}{B\arabic{equation}}
\section{Parker Spiral Lines and Reference Simulations} \label{sec:B}

Parker spiral lines are defined by the following mapping
\begin{equation}
     \frac{dr}{v_{\rm sw}}=\frac{d\phi}{\omega_{s}},
     \label{con:Eq.C.1}
\end{equation}
where $v_{\rm sw}$ is radial solar wind velocity, and $\omega_{s}$ denotes the angular velocity of rotation near the solar equator.
The Parker spiral angle $\phi\left(r\right)$ at heliocentric distance is formulated by
\begin{equation}
     \phi\left(r\right)=\phi_{0}-\frac{\omega_{s}}{v_{\rm sw}}\left(r-r_{0}\right),
     \label{con:Eq.C.2}
\end{equation}
where $\phi_{0},r_{0}$ represent the initial position of Parker spiral lines, usually defined by the location of PSP.
We then derive the Parker ballistic backmapping from PSP in situ measurements to the source surface using Eq.\ref{con:Eq.C.2}
\begin{equation}
     \mathcal{\tilde{P}} \to \mathcal{\tilde{Q}}, \quad 
     \left(r_{\rm psp},\theta_{\rm psp},\phi_{\rm psp}\right) \mapsto \left(R_{ss},\theta_{\rm psp},\phi_{\rm psp}-\frac{\omega_{s}}{v_{\rm sw}}\left(R_{ss}-r_{\rm psp}\right)\right),
     \label{con:Eq.C.3}
\end{equation}
and the magnetic field backmapping is defined as
\begin{equation}
     \begin{cases}
          B_{r}\left(\mathcal{\tilde{P}}\right)=\left(\frac{R_{ss}}{r_{\rm psp}}\right)^{2}B_{r}\left(\mathcal{\tilde{Q}}\right), \\
          B_{\theta}\left(\mathcal{\tilde{P}}\right)=0 \\
          B_{\phi}\left(\mathcal{\tilde{P}}\right)=\left(\frac{\omega_{s}}{v_{\rm sw}}\right)\left(r_{\rm psp}-R_{ss}\right)\sin\left(\theta\right)B_{r}\left(\mathcal{\tilde{P}}\right). \\ 
     \end{cases} 
     \label{con:Eq.C.4}
\end{equation}

To illustrate the two inputs to the objective functional, we run reference PFSS extrapolations with $R_{ss}=2.0R_{s}$ and $2.5R_{s}$ for PSP Encounters 1-19 and extract the corresponding $B_{r}^{\rm k}$ time series. Figure S3 in Supporting Information shows these results for the 19 Carrington rotations. Red and blue markers denote the measured radial magnetic field, and black and orange markers denote the two fixed $R_{ss}$ reference solutions. Figure S1 in Supporting Information shows the corresponding source surface field distributions.

We use Table S2 to compare the magnetic field scaling factor, the normalized objective functional, the supplementary rescaled objective values reported in the table, and the polarity prediction accuracy for the reference solutions. Because the magnetic field amplitude changes strongly from one encounter to another, we normalize the objective functional separately for each encounter using the encounter mean $\mu^{\rm k}$ and standard deviation $\sigma^{\rm k}$. This normalization allows a consistent comparison across PSP encounters.

The scaling factors satisfy $\alpha^{\rm k}>1$ throughout Table S2, which reflects the familiar underestimation of open flux by PFSS extrapolation. Lowering $R_{ss}$ increases the amount of magnetic flux treated as open under the boundary condition used in the PFSS extrapolation, so the $R_{ss}=2.0R_{s}$ reference case generally gives smaller scaling factors than the $R_{ss}=2.5R_{s}$ case. The scaling factors also tend to decrease from the early encounters to the later ones, which is consistent with the improvement seen during the ascending phase in both PSP and ACE comparisons.

Table S3 shows the corresponding ACE comparisons. The polarity prediction accuracy is generally higher during the ascending phase, and the objective functional shows the same broad trend. Because polarity prediction accuracy is not continuously differentiable, we do not use it as the single objective in Algorithm \ref{alg:rss_optimization}. We instead use the normalized objective functional as the optimization target and reserve the scaling factor, white light image comparisons, and coronal hole area fractions for evaluation and interpretation.

\setcounter{figure}{0}
\setcounter{table}{0}
\setcounter{equation}{0}
\renewcommand{\thefigure}{C\arabic{figure}}
\renewcommand{\thetable}{C\arabic{table}}
\renewcommand{\theequation}{C\arabic{equation}}
\section{Argument} \label{sec:D}
\subsection{Lemma 2.1}
The proof of lemma \ref{lemma:2.1} is shown below.
\begin{proof}
     Necessity follows directly from integration by parts.
     Taking an arbitrary $\phi \in V\left(\Omega\right)$, multiply the Laplace equation $-\Delta \psi=0$ with $\phi$ and integrate to obtain
     \begin{equation}
          \int_{\Omega}\left(-\Delta\psi\right)\phi \ dx=0,
          \label{con:Eq.7}
     \end{equation}
     where $\left(-\Delta\psi\right)\phi \in L^{1}\left(\Omega\right)$ by Cauchy Schwarz inequality:
     \begin{equation}
          \Vert \left(\Delta\psi\right)\phi \Vert_{L^{1}\left(\Omega\right)}\leq\Vert \Delta\psi \Vert_{L^{2}\left(\Omega\right)}\Vert \phi \Vert_{L^{2}\left(\Omega\right)}.
          \label{con:Eq.8}
     \end{equation}
     Applying integration by parts to \eqref{con:Eq.7} leads to the variational form
     \begin{equation}
         \begin{aligned}
         \int_{\Omega}\left(-\Delta\psi\right)\phi \ dx&=\sum_{i=1}^{3}\int_{\Omega}\left(-\partial_{i}\partial_{i}\psi\right)\phi \ dx\\
         &=\sum_{i=1}^{3}\left(-\int_{\partial\Omega}\left(\gamma_{0}\left(\partial_{i}\psi \right)\cdot n_{i}\right)\gamma_{0}\left(\phi\right)\ d\sigma +\int_{\Omega}\left(\partial_{i}\psi\right)\left(\partial_{i}\phi\right)\ dx\right)\\
         &=-\int_{\Sigma\cup\Gamma}\gamma_{0}\left(\phi\right)\gamma_{0}\left(\nabla\psi \right)\cdot \mathbf{n}\ d\sigma+\int_{\Omega}\left(\nabla\psi\right)\cdot\left(\nabla\phi\right)\ dx\\
         &=-\int_{\Gamma}\gamma_{0}\left(\phi\right)g\ d\sigma+\int_{\Omega}\left(\nabla\psi\right)\cdot\left(\nabla\phi\right)\ dx\\&
         =0,  
         \end{aligned}
         \label{con:Eq.9}
     \end{equation}
     where the index i represents the ith component subspace of $\mathbb{R}^{3}$.
     From \eqref{con:Eq.9}, we prove to the necessity.
     
     Conversely, if there exists $\psi\in V\left(\Omega\right)$ such that
     \begin{equation}
          \int_{\Omega}\left(\nabla \psi \right)\cdot \left(\nabla \phi \right) dx = \int_{\Gamma}g \gamma_{0}\left(\phi\right) d\sigma,\quad \forall \phi\in V\left(\Omega\right).
          \label{con:Eq.10}
     \end{equation}
     Hence, if $\phi\in \mathcal{D}\left(\Omega\right)\coloneqq\left\{v\in V: \partial_{i}v\in L_{loc}^{2}\left(\Omega\right)\right\}$, 
     \begin{equation}
          \begin{aligned}
          \int_{\Omega}\left(\nabla \psi \right)\cdot \left(\nabla \phi \right) dx &= \int_{\Gamma}g \gamma_{0}\left(\phi\right) d\sigma\\
          &=0,\quad \forall \phi\in \mathcal{D}\left(\Omega\right).
          \end{aligned}
          \label{con:Eq.11}
     \end{equation}
     We conclude that $\Delta\psi=0,\ \psi\in V\left(\Omega\right)$.
     Now we prove that $\psi$ satisfies the boundary conditions on $\Gamma$ and $\Sigma$.
     It is evident that the boundary conditions hold on $\Sigma$ due to $\psi \in V\left(\Omega\right)$.
     By integrating the left hand side of \eqref{con:Eq.6} by parts, we obtain:
     \begin{equation}
          \int_{\Gamma}\gamma_{0}\left(\phi\right)\gamma_{0}\left(\nabla\psi \right)\cdot \mathbf{n}\ d\sigma-\int_{\Omega}\left(\Delta\psi\right)\phi dx = \int_{\Gamma}g \gamma_{0}\left(\phi\right) d\sigma,\quad \forall \phi\in V\left(\Omega\right).
          \label{con:Eq.12}
     \end{equation}
     Then $\Delta\psi$ vanishes, simplifying formula \eqref{con:Eq.12} to
     \begin{equation}
          \int_{\Gamma}\gamma_{0}\left(\phi\right)\left(\gamma_{0}\left(\nabla\psi \right)\cdot \mathbf{n}-g\right)\ d\sigma=0, \quad \forall \phi\in V\left(\Omega\right).
          \label{con:Eq.13}
     \end{equation}
     Selecting a suitable $\psi\in V\left(\Omega\right)$ such that $\gamma_{0}\left(\psi\right)=\gamma_{0}\left(\nabla\psi \right)\cdot \mathbf{n}-g$, we obtain
     \begin{equation}
          \int_{\Gamma}\left(\gamma_{0}\left(\nabla\psi \right)\cdot \mathbf{n}-g\right)^2\ d\sigma=0.
          \label{con:Eq.14}
     \end{equation}
     That implies $\partial_{\mathbf{n}}\psi|_{\Gamma}=g$.
\end{proof}

\subsection{The Proof of Theorem 2.1}
The proof of theorem \ref{thm:1} is shown below.
\begin{proof}
     We first demonstrate the $V-\text{elliptic}$ property of $a\left(\cdot,\cdot\right)$ via the Poincar\'{e} inequality
\begin{equation}
     \begin{aligned}
     a\left(\phi,\phi\right)&=\int_{\Omega}|\nabla \psi|^{2}dx\\
     &=\Vert\nabla \psi\Vert_{L^{2}\left(\Omega\right)}^{2}\geq C^{2}\Vert\psi\Vert_{V\left(\Omega\right)}^{2}, \quad \forall \phi\in V\left(\Omega\right).
     \end{aligned}
     \label{con:Eq.18}
\end{equation}
Hence, $a\left(\cdot,\cdot\right)$ is $V-\text{elliptic}$ with coefficient $\alpha\coloneqq C^{-2}$.
The continuity of $a\left(\cdot,\cdot\right)$ follows directly from the Cauchy Schwarz inequality:
\begin{equation}
     \begin{aligned}
          |a\left(\psi,\phi\right)|&=|\int_{\Omega}\left(\nabla \psi\right)\cdot\left(\nabla \phi\right)dx|\\
          &\leq \Vert \nabla\psi\Vert_{L^{2}\left(\Omega\right)}\cdot\Vert \nabla\phi\Vert_{L^{2}\left(\Omega\right)}\\
          &\leq \Vert \nabla\psi\Vert_{V\left(\Omega\right)}\cdot\Vert \nabla\phi\Vert_{V\left(\Omega\right)},\quad \forall \left(\psi,\phi\right)\in V\times V.
     \end{aligned}
     \label{con:Eq.19}
\end{equation}
Similarly, the continuity of $\eta_{g}\left(\phi\right)$ is established by the trace theorem:
\begin{equation}
     \begin{aligned}
          |\eta_{g}\left(\phi\right)|&=|\int_{\Gamma}g\gamma_{0}\left(\phi\right)dx|\\
          &\leq \Vert g\Vert_{L^{2}\left(\Gamma\right)}\cdot\Vert \gamma_{0}\left(\phi\right)\Vert_{L^{2}\left(\Gamma\right)}\\
          &\leq \tilde{C}\Vert g\Vert_{L^{2}\left(\Gamma\right)}\cdot\Vert \phi\Vert_{V\left(\Omega\right)},\quad \forall \phi\in V\left(\Omega\right).
     \end{aligned}
     \label{con:Eq.20}
\end{equation}
Hence, $a\left(\cdot,\cdot\right)$ and $\eta_{g}\left(\cdot\right)$ satisfy the conditions of the Lax-Milgram theorem.
That proves existence and uniqueness of the solution for PFSS extrapolation in $V\left(\Omega\right)$.

\end{proof}

\subsection{The Proof of Lemma 2.2}
The proof of lemma \ref{lem:2.2} is shown below.
\begin{proof}
     The homeomorphic mappings $\varphi_{n}$ satisfy uniform boundedness and equicontinuity, 
     implying $\mathcal{P}$ is weakly compact by the Arzel\`{a}-Ascoli theorem. 
     Hence, any sequence in $\mathcal{P}$ admits a convergent subsequence $\varphi_{n_{k}}\rightarrow\varphi$.
     The convergence of parameterized domains $\varphi_{n}$ follows from the weak compactness of $\mathcal{P}$.
     
     The last two sequences are convergent due to the boundedness of \eqref{con:Eq.23} and estimations on the variational equations.
     The variational equations for two auxiliary problems are
     \begin{equation}
          \text{find $\psi_{D_{n_{k}}}\in V\left(\Omega_{n_{k}}\right)$, such that:}
          \int_{\Omega_{n_{k}}}\nabla \psi_{D_{n_{k}}} \nabla \phi dx - \int_{\Gamma}g \phi d\sigma=0,\ \forall \phi \in V\left(\tilde{U}\right),
          \label{con:Eq.24}
     \end{equation}
     \begin{equation}
          \begin{aligned}
               \text{find $\psi_{R_{n_{k}}}\in H^1\left(\Omega_{n_{k}}\right)$, such that:} 
               \int_{\Omega_{n_{k}}}\nabla \psi_{R_{n_{k}}} \nabla \phi dx &+\beta\int_{\Sigma_{n_{k}}}\psi_{R_{n_{k}}} \phi d\sigma-\int_{\Sigma_{n_{k}}}h \phi d\sigma\\
               &- \int_{\Gamma}g \phi d\sigma=0,\forall \phi \in H^{1}\left(\tilde{U}\right).
          \end{aligned}
          \label{con:Eq.25}
     \end{equation}
     Now we prove the convergence of the sequences $\psi_{D_{n_{k}}}\rightarrow\psi_{D}$ in $V\left(\tilde{U}\right)$ and $\psi_{R_{n_{k}}}\rightarrow\psi_{R}$ in $H^1\left(\tilde{U} \right)$.
     
     For $\psi_{R_{n_{k}}}$, we analyze its boundedness via the elliptic property of the governing equations.
     A standard technique is to select the test function $\phi=\psi_{R_{n_{k}}}$, enabling estimation of the upper bound in the $H^{1}\left(\Omega_{n_{k}}\right)$ norm through the following equation
     \begin{equation}
          \int_{\Omega_{n_{k}}}\nabla \psi_{R_{n_{k}}} \nabla \psi_{R_{n_{k}}} dx +\beta\int_{\Sigma_{n_{k}}}\psi_{R_{n_{k}}}\cdot\psi_{R_{n_{k}}} d\sigma-\int_{\Sigma_{n_{k}}}h \psi_{R_{n_{k}}} d\sigma- \int_{\Gamma}g \psi_{R_{n_{k}}} d\sigma=0.
          \label{con:Eq.26}
     \end{equation}
     The $H^{1}\left(\Omega_{n_{k}}\right)$-norm is defined by equation \eqref{con:Eq.5}.
     Using the elliptic property, we derive the estimate $\exists\ \alpha_{k}>0,\text{such that}$
     \begin{equation}
          \int_{\Omega_{n_{k}}}\nabla \psi_{R_{n_{k}}} \nabla \psi_{R_{n_{k}}} dx +\beta\int_{\Sigma_{n_{k}}}\psi_{R_{n_{k}}}\cdot\psi_{R_{n_{k}}} d\sigma\geq \alpha_{k}\Vert\psi_{R_{n_{k}}}\Vert_{H^1\left(\Omega_{n_{k}}\right)}^{2}, 
          \label{con:Eq.27}
     \end{equation}
     We apply the Cauchy Schwarz inequality to estimate the boundary integrals
     \begin{equation}
          \left|\int_{\Sigma_{n_{k}}}h \psi_{R_{n_{k}}} d\sigma+ \int_{\Gamma}g \psi_{R_{n_{k}}} d\sigma\right|\leq\left|\tilde{U}\right|^{\frac{1}{2}}\left(\Vert\psi_{R_{n_{k}}}\Vert_{L^2\left(\Sigma_{n_{k}}\right)}\underset{\Sigma_{n_{k}}}{\max}{\left|h\right|}+\Vert\psi_{R_{n_{k}}}\Vert_{L^2\left(\Gamma\right)}\underset{\Gamma}{\max}{\left|g\right|}\right),
          \label{con:Eq.28}
     \end{equation}
     where $\left|\tilde{U}\right|$ is the maximum diameter of the domain $\tilde{U}$.
     The compactness of the boundaries $\Sigma_{n_{k}}$ and $\Gamma$ ensures the existence of this maximum.
     Combining equations \eqref{con:Eq.27} and \eqref{con:Eq.28}, we obtain
     \begin{equation}
          \begin{aligned}
               \alpha_{k}\Vert\psi_{R_{n_{k}}}\Vert_{H^1\left(\Omega_{n_{k}}\right)}^{2}&\leq\left|\tilde{U}\right|^{\frac{1}{2}}\left(\Vert\psi_{R_{n_{k}}}\Vert_{L^2\left(\Sigma_{n_{k}}\right)}\underset{\Sigma_{n_{k}}}{\max}{\left|h\right|}+\Vert\psi_{R_{n_{k}}}\Vert_{L^2\left(\Gamma\right)}\underset{\Gamma}{\max}{\left|g\right|}\right)\\
               &\leq \tilde{C}\left(\Vert\psi_{R_{n_{k}}}\Vert_{L^2\left(\Sigma_{n_{k}}\right)}+\Vert\psi_{R_{n_{k}}}\Vert_{L^2\left(\Gamma\right)}\right),
          \end{aligned}
          \label{con:Eq.29}
     \end{equation}
     where $\tilde{C}=\max\{\left|\tilde{U}\right|^{\frac{1}{2}}\underset{\Sigma_{n_{k}}}{\max}{\left|h\right|},\left|\tilde{U}\right|^{\frac{1}{2}}\underset{\Gamma}{\max}{\left|g\right|}\}$ and the right hand side of inequality \eqref{con:Eq.29} can be further raised using the trace theorem:
     \begin{equation}
          \begin{aligned}
               \Vert\psi_{R_{n_{k}}}\Vert_{L^2\left(\Sigma_{n_{k}}\right)}\leq C_{1}\Vert\psi_{R_{n_{k}}}\Vert_{H^1\left(\Omega_{n_{k}}\right)}\leq \tilde{C_{1}}\Vert\psi_{R_{n_{k}}}\Vert_{H^1\left(\tilde{U}\right)},\\
               \Vert\psi_{R_{n_{k}}}\Vert_{L^2\left(\Gamma\right)}\leq C_{2}\Vert\psi_{R_{n_{k}}}\Vert_{H^1\left(\Omega_{n_{k}}\right)}\leq \tilde{C_{2}}\Vert\psi_{R_{n_{k}}}\Vert_{H^1\left(\tilde{U}\right)},
          \end{aligned}
          \label{con:Eq.30}
     \end{equation}
     Then, the norm is formulated by the domain extension with the constants $E_{1},E_{2}$ to derive the following equivalent theorem:
     \begin{equation}
          E_{1}\Vert\psi\Vert_{H^1\left(\tilde{U}\right)} \leq \Vert\psi\Vert_{H^1\left(\Omega_{n_{k}}\right)} \leq E_{2}\Vert\psi\Vert_{H^1\left(\tilde{U}\right)}.
          \label{con:Eq.31}
     \end{equation}
     By the formula \eqref{con:Eq.29}, \eqref{con:Eq.30} and \eqref{con:Eq.31}, we conclude the sequence $\psi_{n_{k}}$ is bounded.
     \begin{equation}
          \begin{aligned}
               \alpha_{k}\Vert\psi\Vert^{2}_{H^1\left(\Omega_{n_{k}}\right)}&\leq \tilde{C}\left(\Vert\psi_{R_{n_{k}}}\Vert_{L^2\left(\Sigma_{n_{k}}\right)}+\Vert\psi_{R_{n_{k}}}\Vert_{L^2\left(\Gamma\right)}\right)\\
               &\leq \tilde{C}\max\{\tilde{C_{1}},\tilde{C_{2}}\}\Vert\psi_{n_{k}}\Vert_{H^1\left(\tilde{U}\right)}\\
               &\leq \tilde{C}\max\{\tilde{C_{1}},\tilde{C_{2}}\}E^{-1}_{1}\Vert\psi\Vert_{H^1\left(\Omega_{n_{k}}\right)},
          \end{aligned}
          \label{con:Eq.32}
     \end{equation}
     where we denote the extension factor by $c=\left(\tilde{C}\max\{\tilde{C_{1}},\tilde{C_{2}}\}\right)\cdot\left(E_{1}\alpha_{k}\right)^{-1}$.
     Then, the boundedness of sequence $\left\{\psi_{n_{k}}\right\}$ is proven:
     \begin{equation}
          \Vert\psi_{n_{k}}\Vert_{H^1\left(\tilde{U}\right)}\leq c\cdot E_{1}.
          \label{con:Eq.33}
     \end{equation}
     Hence, there exists a unique limit point $\psi_{R}$ for this bounded convergent sequence in $H^{1}\left(\tilde{U}\right)$.
     The above proof applies analogously to the convergence $\psi_{D_{n_{k}}}\rightarrow\psi_{D}$ in $V\left(\tilde{U}\right)$.
     Next, we verify that the limit points $\psi_{D}$ and $\psi_{R}$ satisfy the variational equations \eqref{con:Eq.24} and \eqref{con:Eq.25} in the domain $\Omega\left(\Sigma\left(\varphi\right)\right)$, respectively.
     
     Considering the $\psi_{R}\left(\varphi\right)=\psi_{R}|_{\Omega\left(\Sigma\left(\varphi\right)\right)}$, the variational equation \eqref{con:Eq.25} can be expressed as
     \begin{equation}
          \int_{\Omega\left(\Sigma\left(\varphi\right)\right)}\nabla \psi_{R} \nabla \phi dx +\beta\int_{\Sigma\left(\varphi\right)}\psi_{R} \phi d\sigma-\int_{\Sigma\left(\varphi\right)}h \phi d\sigma- \int_{\Gamma}g \phi d\sigma=0,
          \label{con:Eq.34}
     \end{equation}
     where $\phi$ is the test function in $H^{1}\left(\Omega\left(\Sigma\left(\varphi\right)\right)\right)$.
     
     For the subsequence $\psi_{n_{k}}$ of the solution to Eq.\eqref{con:Eq.25}, the variational equation is
     \begin{equation}
          \int_{\Omega_{n_{k}}}\nabla \psi_{R_{n_{k}}} \nabla \phi dx +\beta\int_{\Sigma_{n_{k}}}\psi_{R_{n_{k}}} \phi d\sigma-\int_{\Sigma_{n_{k}}}h \phi d\sigma- \int_{\Gamma}g \phi d\sigma=0.
          \label{con:Eq.35}
     \end{equation}
     We consider the difference between \eqref{con:Eq.34} and \eqref{con:Eq.35},
     where the test function $\phi$ in $H^1\left(\tilde{U}\right)$ is limited to $H^{1}\left(\Omega\left(\Sigma\left(\varphi\right)\right)\right)$.
     Taking the limit $k\rightarrow\infty$,
     the difference of two variational equations can be formulated as follows:
     \begin{equation}
          \begin{aligned}
               D_{1}&=\left|\int_{\Omega\left(\Sigma\left(\varphi\right)\right)}\nabla \psi_{R} \nabla \phi dx-\int_{\Omega_{n_{k}}}\nabla \psi_{R_{n_{k}}} \nabla \phi dx\right|,\\
               D_{2}&=\beta\left|\int_{\Sigma\left(\varphi\right)}\psi_{R} \phi d\sigma-\int_{\Sigma_{n_{k}}}\psi_{R_{n_{k}}} \phi d\sigma\right|,\\
               D_{3}&=\left|\int_{\Sigma\left(\varphi\right)}h \phi d\sigma-\int_{\Sigma_{n_{k}}}h \phi d\sigma\right|,\\
               D_{4}&=\left|\int_{\Gamma}g \phi d\sigma-\int_{\Gamma}g \phi d\sigma\right|,\\
          \end{aligned}
          \label{con:Eq.36}
     \end{equation}
     where $D_{4}$ is independent of the index $k$, hence $D_{4}=0$. 
    
     Considering the difference $D_{1}$, 
     we extend the integral over the domain $\tilde{U}$ using the characteristic function $\zeta$
     \begin{equation}
          \begin{aligned}
               D_{1}&=\left|\int_{\Omega\left(\Sigma\left(\varphi\right)\right)}\nabla \psi_{R} \nabla \phi dx-\int_{\Omega_{n_{k}}}\nabla \psi_{R_{n_{k}}} \nabla \phi dx\right|\\
               &=\left|\int_{\tilde{U}}\zeta_{\Omega\left(\varphi\right)}\nabla \tilde{\psi}_{R} \nabla \phi dx-\int_{\tilde{U}}\zeta_{\Omega_{k}}\nabla \tilde{\psi}_{R_{n_{k}}} \nabla \phi dx\right|\\
               &=\left|\int_{\tilde{U}}\zeta_{\Omega\left(\varphi\right)}\left(\nabla \tilde{\psi}_{R}-\nabla \tilde{\psi}_{R_{n_{k}}}\right)\nabla \phi dx+\int_{\tilde{U}}\left(\zeta_{\Omega\left(\varphi\right)}-\zeta_{\Omega_{k}}\right)\nabla \tilde{\psi}_{R_{n_{k}}} \nabla \phi dx\right|\\
               &\leq \left|\int_{\tilde{U}}\zeta_{\Omega\left(\varphi\right)}\left(\nabla \tilde{\psi}_{R}-\nabla \tilde{\psi}_{R_{n_{k}}}\right)\nabla \phi dx \right|+\left|\int_{\tilde{U}}\left(\zeta_{\Omega\left(\varphi\right)}-\zeta_{\Omega_{k}}\right)\nabla \tilde{\psi}_{R_{n_{k}}} \nabla \phi dx\right|,
          \end{aligned}
          \label{con:Eq.37}
     \end{equation}
     where $\tilde{\psi}_{R}$, $\tilde{\psi}_{R_{n_{k}}}$ are the zero extensions of $\psi_{R}$, $\psi_{R_{n_{k}}}$ in $\tilde{U}$, respectively.
     As $k\rightarrow\infty$, the first term on the right hand side converges to zero,
     as demonstrated by the convergence of the sequence $\psi_{n_{k}}$ in $H^1\left({\tilde{U}}\right)$.
     The convergence of the characteristic functions $\zeta_{k}$ corresponds the convergence of the domain sequence parameterized by $\varphi_{n_{k}}$ for the second term.
     
     Considering the difference $D_{2}$, we first perform a parameter transformation using $y$ along the boundary $\Sigma$ and estimate the upper bound via triangle inequality:
     \begin{equation}
          \begin{aligned}
               D_{2}&=\beta\left|\int_{\Sigma\left(\varphi\right)}\psi_{R} \phi d\sigma-\int_{\Sigma_{n_{k}}}\psi_{R_{n_{k}}} \phi d\sigma\right|\\
               &=\beta\left|\int_{\mathbb{S}^{2}}\left(\psi_{R}\circ \varphi\left(y\right)\right)\left(\phi\circ \varphi\left(y\right)\right) J_{\varphi}\left(y\right)-\left(\psi_{R_{n_{k}}}\circ \varphi_{n_{k}}\left(y\right)\right)\left(\phi\circ \varphi_{n_{k}}\left(y\right)\right) J_{\varphi_{k}}\left(y\right)d\sigma\left(y\right)\right|\\
               &\leq \beta\left|\int_{\mathbb{S}^{2}}\left(\psi_{R}\circ \varphi\right)\left(\phi\circ \varphi\right)\left(J_{\varphi}-J_{\varphi_{k}}\right)d\sigma\right|+\beta\left|\int_{\mathbb{S}^{2}}\left(\psi_{R}\circ \varphi\right)\left(\phi\circ \varphi-\phi\circ \varphi_{n_{k}}\right) J_{\varphi_{k}}d\sigma\right|\\
               & + \beta\left|\int_{\mathbb{S}^{2}}\left(\psi_{R}\circ \varphi-\psi_{R_{n_{k}}}\circ \varphi\right)\left(\phi\circ \varphi_{n_{k}}\right) J_{\varphi_{k}}d\sigma\right|\\
               &+\beta\left|\int_{\mathbb{S}^{2}}\left(\psi_{R_{n_{k}}}\circ \varphi-\psi_{R_{n_{k}}}\circ \varphi_{R_{n_{k}}}\right)\left(\phi\circ \varphi_{n_{k}}\right) J_{\varphi_{k}}d\sigma\right|,
          \end{aligned}
          \label{con:Eq.38}
     \end{equation}
     where $J_{\varphi}$ is the Jacobian of parameter transformation represented by $\varphi=r\left(y\right)y,\ y\in\mathbb{S}^2$,
     then $J_{\varphi}=r\cdot\left(r^2+\nabla r\cdot\nabla r\right)^{\frac{1}{2}}$.
     For $\forall\ \varphi\in\mathcal{P}$, there exist positive constants $c_{1},c_{2}$ such that $c_{1}\leq \left|det\left(J_{\varphi}\right)\right|\leq c_{2}$.
     Next, we estimate each term on the right side hand using the Cauchy Schwarz inequality:
     \begin{equation}
          \begin{aligned}
               \text{term 1}&=\beta\left|\int_{\mathbb{S}^{2}}\left(\psi_{R}\circ \varphi\right)\left(\phi\circ \varphi\right)\left(J_{\varphi}-J_{\varphi_{k}}\right)d\sigma\right|\\
               &=\beta\left|\int_{\mathbb{S}^{2}}\left(\psi_{R}\circ \varphi\right)\left(\phi\circ \varphi\right)J_{\varphi}\frac{\left(J_{\varphi}-J_{\varphi_{k}}\right)}{J_{\varphi}}d\sigma\right|\\
               &\leq \beta\  \underset{\mathbb{S}^2}{\sup}\left|J_{\varphi}-J_{\varphi_{k}}\right|\cdot c_{1}^{-1}\left|\int_{\mathbb{S}^{2}}\left(\psi_{R}\circ \varphi\right)\left(\phi\circ \varphi\right)J_{\varphi}d\sigma\right|\\
               &\leq \beta\  \underset{\mathbb{S}^2}{\sup}\left|J_{\varphi}-J_{\varphi_{k}}\right|\cdot c_{1}^{-1}\Vert\psi_{R}\Vert_{L^{2}\left(\Sigma\left(\varphi\right)\right)}\Vert\phi\Vert_{L^{2}\left(\Sigma\left(\varphi\right)\right)},\\
               \text{term 2}&=\beta\left|\int_{\mathbb{S}^{2}}\left(\psi_{R}\circ \varphi\right)\left(\phi\circ \varphi-\phi\circ \varphi_{n_{k}}\right) J_{\varphi_{k}}d\sigma\right|\\
               &=\beta\left|\int_{\mathbb{S}^{2}}\left(\psi_{R}\circ \varphi\right)\left(\phi\circ \varphi-\phi\circ \varphi_{n_{k}}\right) J_{\varphi_{k}}\cdot\sqrt{\frac{J_{\varphi}}{J_{\varphi}}}d\sigma\right|\\
               &\leq \beta\ \left|\int_{\mathbb{S}^{2}}\left(\psi_{R}\circ \varphi\right)^{2}J_{\varphi_{k}}^{2}\cdot\frac{J_{\varphi}}{J_{\varphi}} d\sigma\right|^{\frac{1}{2}}\cdot\Vert\left(\phi\circ \varphi-\phi\circ \varphi_{n_{k}}\right)\Vert_{L^{2}\left(\mathbb{S}^{2}\right)}\\
               &\leq \beta\  c_{2}\cdot c_{1}^{-\frac{1}{2}}\Vert\psi_{R}\Vert_{L_{\Sigma\left(\varphi\right)}^{2}}\Vert\left(\phi\circ \varphi-\phi\circ \varphi_{n_{k}}\right)\Vert_{L^{2}\left(\mathbb{S}^{2}\right)},\\
               \text{term 3}&=\beta\left|\int_{\mathbb{S}^{2}}\left(\psi_{R}\circ \varphi-\psi_{R_{n_{k}}}\circ \varphi\right)\left(\phi\circ \varphi_{n_{k}}\right) J_{\varphi_{k}}d\sigma\right|\\
               &\leq \beta\ c_{2}^{\frac{1}{2}}\cdot c_{1}^{-\frac{1}{2}}\Vert\psi_{R}-\psi_{R_{n_{k}}}\Vert_{L^{2}\left(\Sigma\left(\varphi\right)\right)} \Vert\phi\Vert_{L^{2}\left(\Sigma_{k}\right)},\\
               \text{term 4}&=\beta\left|\int_{\mathbb{S}^{2}}\left(\psi_{R_{n_{k}}}\circ \varphi-\psi_{R_{n_{k}}}\circ \varphi_{R_{n_{k}}}\right)\left(\phi\circ \varphi_{n_{k}}\right) J_{\varphi_{k}}d\sigma\right|\\
               &\leq \beta\ c_{2}^{\frac{1}{2}}\Vert\psi_{R_{n_{k}}}\circ \varphi-\psi_{R_{n_{k}}}\circ \varphi_{R_{n_{k}}}\Vert_{L^{2}\left(\mathbb{S}^{2}\right)} \Vert\phi\Vert_{L^{2}\left(\Sigma_{k}\right)},
          \end{aligned}
          \label{con:Eq.39}
     \end{equation}
     where $\forall\ \varphi\in\mathcal{P},\ J_{\varphi}\neq 0$ by the definition of a regular parameter transformation. 
     The convergence of each term is evident from the sequences $J_{\varphi_{k}}$ and $\psi_{R_{n_{k}}}$ in $H^{1}\left(\tilde{U}\right)$.
     Hence, $D_{2}$ converges to $0$ as $k\rightarrow\infty$.
     
     Considering the difference $D_{3}$, we apply the same parameter transformation method.
     \begin{equation}
          \begin{aligned}
               D_{3}&=\left|\int_{\Sigma\left(\varphi\right)}h \phi d\sigma-\int_{\Sigma_{n_{k}}}h \phi d\sigma\right|,\\
               &= \left|\int_{\mathbb{S}^{2}}\left(h\circ \varphi\right)\left(\phi\circ \varphi\right)J_{\varphi} d\sigma-\int_{\mathbb{S}^{2}}\left(h\circ \varphi_{n_{k}} \right)\left(\phi\circ \varphi_{n_{k}}\right)J_{\varphi_{{k}}} d\sigma\right|,\\
               &\leq \left|\int_{\mathbb{S}^{2}}\left(h\circ \varphi\right)\left(\phi\circ \varphi\right)\left(J_{\varphi}-J_{\varphi_{k}}\right) d\sigma\right|+\left|\int_{\mathbb{S}^{2}}\left(h\circ \varphi-h\circ \varphi_{n_{k}}\right)\left(\phi\circ \varphi\right)J_{\varphi_{{k}}} d\sigma\right|\\
               &+\left|\int_{\mathbb{S}^{2}}\left(h\circ \varphi_{n_{k}}\right)\left(\phi\circ \varphi-\phi\circ \varphi_{n_{k}}\right)J_{\varphi_{k}}d\sigma\right|+\left|\int_{\mathbb{S}^{2}}\left(h\circ \varphi_{n_{k}}\right)\left(\phi\circ \varphi_{n_{k}}-\phi\circ \varphi_{n_{k}}\right)J_{\varphi_{k}}d\sigma\right|,
          \end{aligned}
          \label{con:Eq.40}
     \end{equation}
     where the last term on the right hand side is equal to $0$.
     As in the analysis of $D_{2}$, we infer that $D_{3}$ converges to $0$ as $k\rightarrow\infty$.
     The convergence of sequence $\psi_{D}$ is proven in the same way.

     Hence, equation \eqref{con:Eq.35} converges to the equation \eqref{con:Eq.34} as $k\rightarrow\infty$.
     This proves the compactness of set $\mathcal{M}$ in $\mathcal{P} \times V\left(\tilde{U}\right) \times H^{1}\left(\tilde{U}\right)$.
\end{proof}

\end{sloppypar}
\clearpage

\end{document}